\documentclass[11pt,a4paper]{article}

\usepackage{jcappub}
\pdfoutput=1
\usepackage{multirow,relsize,slashed,textpos}
\usepackage{amssymb}
\usepackage{amsbsy}
\usepackage{amsfonts}
\usepackage{amsmath}
\usepackage{graphicx}
\usepackage{amssymb}
\usepackage{xcolor}
\usepackage{slashed}  
\usepackage{setspace}
\usepackage{amstext}  
\usepackage{wasysym}
\usepackage{mathrsfs}
\usepackage{mdframed} 
\usepackage{booktabs} 
\usepackage{multirow}
\usepackage{subfig}
\usepackage{placeins}
\usepackage{array,mathtools}
\usepackage{braket} 
\usepackage{physics} 
\usepackage{bm} 
\newcolumntype{L}[1]{>{\raggedright\arraybackslash}p{#1}}
\usepackage{accents} 
\usepackage{enumerate}

\newcommand{\CNB}{{\rm C}\nu {\rm B}}

\DeclareMathAlphabet{\mathpzc}{OT1}{pzc}{m}{it}

\newcommand{\eV}{{\rm\ eV}}
\newcommand{\g}{{\rm\ g}}
\newcommand{\keV}{{\rm\ keV}}

\newcommand{\n}{{neutrino}}
\newcommand{\ns}{{neutrinos}}

\graphicspath{{Figures/}}

\title{Dirac and Majorana neutrino signatures of primordial black holes}

\author[a]{Cecilia Lunardini}
\author[b,c,d]{and Yuber F. Perez-Gonzalez}

\affiliation[a]{Department of Physics, Arizona State University, 450 E. Tyler Mall, Tempe, AZ 85287-1504 USA}
\affiliation[b]{Theoretical Physics Department, Fermi National Accelerator Laboratory, P.O. Box 500, Batavia, IL 60510, USA}
\affiliation[c]{Department of Physics \& Astronomy, Northwestern University, Evanston, IL 60208, USA}
\affiliation[d]{Colegio de F\'isica Fundamental e Interdisciplinaria de las Am\'ericas (COFI), 254 Norzagaray street, San Juan, Puerto Rico 00901.} 

\emailAdd{cecilia.lunardini@asu.edu}
\emailAdd{yfperezg@northwestern.edu}

\preprint{FERMILAB-PUB-19-521-T, NUHEP-TH/19-14}

\abstract{

We study Primordial Black Holes (PBHs) as sources of massive neutrinos via Hawking radiation. Under the hypothesis that black holes emit neutrino mass eigenstates, we describe quantitatively how the PBH evolution and lifetime is affected by the mass and fermionic --- Dirac or Majorana --- nature of neutrinos. In the case of Dirac neutrinos, PBHs radiate right-handed and left-handed neutrinos in equal amounts, thus possibly increasing the effective number of neutrino species, $N_{\rm eff}$. Assuming an initially monochromatic PBH mass spectrum, with the initial mass $M_i$ related to the particle horizon mass, and considering the current constraint on $N_{\rm eff}$, we derive a  bound on the initial PBH fraction $\beta^\prime$ in the interval $4.3\times 10^7\ {\rm g}\lesssim M_i \lesssim 10^9$ g. Future measurements of $N_{\rm eff}$ may be able to constraint the initial fraction for black hole  masses as low as 1 g. If an excess in $N_{\rm eff}$ is found, PBHs with Dirac neutrinos could provide a minimal explanation of it. For example, for $10^7\ {\rm g} \lesssim M_i\lesssim 10^9$ g and $\beta^\prime \gtrsim 10^{-13}$, an excess radiation at the level of $0.2\lesssim \Delta N_{\rm eff}\lesssim  0.37$ is produced, which can alleviate the tension of the Hubble parameter measurements. Finally, we obtain the  diffuse flux of right-helical neutrinos from PBHs at the Earth, and show that their detection in a PTOLEMY-like detector (using neutrino capture on tritium) would be difficult.
}

\begin{document}

\maketitle
\flushbottom

\section{Introduction}

The existence of black holes (BHs) in the Universe is now well established. The 2016 discovery of gravitational waves from the merger of stellar-mass BHs  \cite{Abbott:2016blz} is a direct evidence of it, and has stimulated a wide range of studies of BH phenomenology. In this context, various mechanisms to explain the origins of BHs in astrophysics and cosmology have been considered. One possibility is that BHs might be produced in the early Universe shortly after inflation, as a result of the gravitational collapse of density fluctuations \cite{Hawking:1971ei,Carr:1974nx,Carr:2005zd,Khlopov:2008qy,Carr:2009jm}. These {\it primordial black holes} (PBHs)  can have masses exceeding the Planck mass, and their Schwarzschild radius can be small enough for quantum effects to be important. They could constitute (part of) the Dark Matter (DM); a possibility that has gained attention recently \cite{Ali-Haimoud:2016mbv,Bird:2016dcv,Carr:2016drx,Inomata:2017okj,Sasaki:2018dmp}.

As Hawking demonstrated in early seminal papers \cite{Hawking:1974rv,Hawking:1974sw}, black holes evolve with time -- and eventually vanish out of existence -- by losing mass via particle radiation. For PBHs, this evaporation process can have observable effects, which allow to place constraints on PBH models and parameters \cite{Carr:2009jm,Carr:2017jsz,Lennon:2017tqq}. Interestingly, a number of phenomenological effects of PBH evaporation are related to their \n\ emission, which can be primary (direct emission as Hawking radiation) or secondary (via the decay of leptons and hadrons) \cite{Carr:1976zz, MacGibbon:1990zk, MacGibbon:1991tj, Halzen:1995hu, Bugaev:2000bz,  Bugaev:2002yt, Bambeck:2005bz, Carr:2009jm}. 

Constraints from neutrino emission have focused on PBHs with masses $M_i\gtrsim 10^9$ g \cite{Carr:1976zz,Halzen:1995hu,Bugaev:2000bz, Bugaev:2002yt, Dave:2019epr}. Limits from atmospheric and solar \n\ experiments \cite{Bugaev:2000bz, Bugaev:2002yt} and from the search of astrophysical $\bar{\nu}_e$ at SuperKamiokande \cite{Malek:2002ns} have been considered, however they are weaker than BBN or $\gamma$-ray limits \cite{Carr:2009jm}. The kilometer-scale detector IceCube is sensitive to the high-energy neutrinos emitted in the last $\sim 10^3$ s before the PBH disappearance \cite{Dave:2019epr}, and could provide constraints on PBH parameters at a level similar to $\gamma$-ray bounds \cite{Halzen:1995hu}. Very recently, the production of light non-interacting states (such as sterile \ns) via the Hawking radiation in a possible BH dominated era has been analysed, finding that it could alleviate the tension between the measurements of the Hubble parameter \cite{Hooper:2019gtx}.

While most literature so far has considered \ns\ as massless, initial conceptual studies have pointed out the potential importance of including \n\ mass effects. As early as 1970's, it was pointed out that the helicity suppression present in the weak interactions (see, e.g. \cite{Kayser:1982br,Nieves:1981zt,Kayser:1983wm, Menon:2008wa}) is absent in the Hawking radiation \cite{Unruh:1976fm}. Later, effects of \n\ masses on the BH evaporation have been considered qualitatively \cite{Bambeck:2005bz,Ukwatta:2015iba}. These early works left open the question of how these effects could possibly impact the diverse PBH phenomenology and cosmology. The time is now mature to address this question, in the light of the greatly advanced picture we now have of neutrino masses and mixing (e.g., \cite{Kajita:2016cak, McDonald:2016ixn,Esteban:2018azc}), and also in the context of the renewed attention for PBH physics. In this paper, assuming that the PBHs were formed after inflation, we want to address carefully the imprint of neutrino masses and fermionic nature on the PBH phenomenology and possible cosmological implications. Furthermore we will show that the case of Dirac \ns\ provides a minimal realization of the scenario of Ref. \cite{Hooper:2019gtx}, with PBHs radiating light, non-interacting right-handed neutrinos. 

This manuscript is organized as follows. In Sec. \ref{sec:primnuem}, we discuss the general neutrino emission from Schwarschild BHs and the impact on their evaporation of nonzero neutrino masses and of the two possible fermionic natures, Dirac or Majorana. Supposing the minimal extension of the Standard Model (SM) to accommodate neutrinos as Dirac particles, in Sec. \ref{sec:const} we derive a constraint on the initial PBH fraction, given the possibility of emitting additional radiation in form of right-handed neutrinos. The implications of Dirac \ns\ for reconciling measurements of the Hubble parameter are briefly presented as well.  In Sec. \ref{sec:diffflux}, the diffuse neutrino flux at Earth and its detectability are discussed.
Finally, in Sec. \ref{sec:conc} we draw our conclusions. We will consider natural units in which $c=\hbar=k_{B}=1$ throughout this article. 

\section{Primary neutrino emission from primordial black holes}\label{sec:primnuem}

In this section we summarize the main features of the emission of {\it massive} \ns\ from PBHs. We will mainly focus on the primary \n\ emission, where effects of the \n\ mass can be strong. Due to the emphasis on mass effects, secondary \n\ production from the decay of leptons or hadrons, where mass effects are negligible, will not be discussed. We will also limit the discussion to Schwarzschild black holes (i.e., black holes with zero charge and zero angular momentum). This is justified because possible non-zero charge and/or angular momentum initially present in a BH would evaporate much faster than the mass, ultimately leading to a Schwarzschild BH \cite{Hawking:1974rv,Hawking:1974sw,Page:1976df,Page:1976ki,Page:1977um} .

\subsection{Massive neutrinos from Schwarzschild black holes}
\label{sub:emission}

In the SM neutrinos are massless Weyl particles with three weak interaction eigenstates (flavors): $\nu_{\alpha}$ ($\alpha=e,\mu,\tau$). It has been established experimentally \citep{Kajita:2016cak,McDonald:2016ixn} that \ns\ undergo \emph{flavor oscillations}, that is, a \n\ initially produced in a given flavor state, can later (after propagation in vacuum or in a medium) be detected with a different flavor. 
This phenomenon arises as consequence of two important features: (i) neutrinos are massive particles and (ii) the mass and flavor bases do not coincide, but rather, the flavor eigenstates  are coherent superpositions of the mass eigenstates $\nu_a$ ($a=1,2,3$), with coefficients given by the Pontecorvo-Maki-Nakagawa-Sakata (PMNS) mixing matrix, $U$: $\nu_\alpha = \sum_a U_{\alpha a}\nu_a$. The massive neutrino fields $\nu_a$ are the truly fundamental quantities, since they have definite kinematic\footnote{Flavor eigenstates, for example, do not have definite masses~\cite{Giunti:2003dg}.} and asymptotic properties, and can be canonically quantized in a standard fashion~\cite{Giunti:2003dg,Ho:2012yja}. All weak processes can be ge\-ne\-ra\-li\-zed consistently to include massive neutrino fields with mixing~\cite{Shrock:1980vy,Shrock:1980ct,Shrock:1981wq}. For instance, the charged-current leptonic lagrangian is written in terms of massive neutrino fields as:
\begin{align}\label{eq:CCLag} 
	\mathscr{L}_{\rm CC}^\ell=-\frac{g}{\sqrt{2}}U_{\alpha a}^*\overline{\nu_{a L}}\gamma^\mu \ell_{\alpha L} W_\mu+{\rm h.c.}\,
\end{align}
with $g$ the gauge coupling, $\ell_\alpha$ the charged lepton fields, and $W_\mu$ the W boson gauge field.  Therefore, a weak interaction process, where a \n\ flavor eigenstate is emitted, can be viewed as a process where the all three neutrino mass eigenstates are produced simultaneously, in a coherent linear superposition, with weights that are determined by the relevant PMNS matrix elements. 

Being massive and neutral fundamental fermions, neutrinos could be their own antiparticles. Therefore, there are two possibilities:
\begin{enumerate}
\item[(i)] \emph{Majorana neutrinos}, where lepton number is violated, and neutrinos and antineutrinos coincide.
In this case, for each mass eigenstate $a$, we have two degrees of freedom,  neutrinos with left (LH) and right (RH) helicities,
\begin{align*}
	\nu_{aL},\ \nu_{aR}~.
\end{align*}
Here, for sake of simplicity, we do not consider the origin of neutrino masses in detail to avoid extra assumptions in the present discussion. Nevertheless, we note in passing that heavy right-handed Majorana neutrinos like those appearing in the Seesaw mechanism \cite{Mohapatra:1979ia,GellMann:1980vs,Mohapatra:1980yp,Schechter:1980gr,Lazarides:1980nt,Mohapatra:1986bd,Foot:1988aq} could also be present. 
\noindent
\item[(ii)] \emph{Dirac neutrinos}, where lepton number can be conserved, and neutrinos ($\nu$) and antineutrinos ($\bar \nu$) are distinct degrees of freedom. Hence, four independent degrees  of freedom  (per mass eigenstate) exist:
\begin{align*}
	\nu_{aL},\ \nu_{aR},\ \overline{\nu_{aL}},\ \overline{\nu_{aR}}.
\end{align*}
Note that the RH and LH states have the same mass in the case of Dirac neutrinos within the minimal extension of the SM~\cite{Balantekin:2018azf}, which is our scenario of choice here. Moreover, the RH neutrino and LH antineutrino states do not couple to the weak interaction, only via Yukawa couplings. Their production in weak processes is strongly suppressed
by a helicity factor, so they are effectively sterile.
\end{enumerate}

Let us now summarize  -- in a brief and incomplete manner, see ~\cite{Hawking:1974rv,Hawking:1974sw,Traschen:1999zr} for a dedicated  treatment -- the physics of \n\ production via Hawking radiation. The Hawking effect is a consequence of the ambiguity of the concept of particle in a curved spacetime. To fix the ideas, let us assume that a BH is formed due to a gravitational collapse event. In the region of spacetime $\mathscr{J^-}$, defined as the far past causally connected (``past null infinity") to the BH formation,  we can choose a set of orthonormal solutions of the field equations for a given spin; these solutions are taken such that it is possible to define modes with positive frequency. Under these conditions, the associated quantum field operator can be expanded in terms of creation and annihilation operators,  $\{{\rm a}_i, {\rm a}_i^\dagger\}$. Let us assume total absence of particles in the past null infinity, that is
\begin{align}
    {\rm a}_i\ket{0}_{\rm in} = 0\, ,
\end{align}
with $\ket{0}_{\rm in}$ being the quantum vacuum state.

After the BH formation, in a region in the far future casually accessible (``future null infinity", $\mathscr{J^+}$), the field operator can be expanded in a different basis, called the \emph{out}-basis.
There will be a different set of annihilation and creation operators, $\{{\rm b}_i,{\rm b}_i^\dagger\}$, of outgoing modes with positive frequency. The two sets of operators at the past and future null infinities are related by a Bogoliubov transformation
\begin{align*}
	{\rm b}_i=\sum_j\alpha_{ij}{\rm a}_j+\beta_{ij}{\rm a}_j^\dag,
\end{align*} 
where the coefficients $\beta_{ij}$ are known as Bogoliubov coefficients. We can see that the state $\ket{0}_{\rm in}$ is not a vacuum state for an observer in the future null infinity $\mathscr{J^+}$, that is, the expectation value of the number operator is nonzero for such observer
\begin{align}
    _{\rm in}\langle 0|{\rm b}_i^\dagger {\rm b}_i|0\rangle_{\rm in}= \sum_j |\beta_{ij}|^2.
\end{align}
After obtaining the explicit form of the coefficients $\beta_{ij}$ \cite{Hawking:1974rv,Hawking:1974sw}, one finds that the expectation value of the number operator on $\mathscr{J^+}$ 
follows a thermal spectrum with a temperature related to the Schwarzschild BH mass $M$ as
\begin{align} \label{eq:TBH}
    T_{\rm BH}=\frac{1}{8\pi G M}\approx 1.06 \left(\frac{10^{13}\ {\rm g}}{M}\right)\ {\rm GeV}~,
\end{align}
\cite{Hawking:1974rv,Hawking:1974sw} (here $G$ is the gravitational coupling). 
Thus, a flux of out-going particles is produced due to the gravitational disturbance created by the collapse \cite{birrell_davies_1982}.

How does the Hawking production mechanism apply to massive neutrinos? Surprisingly, the question has never been discussed specifically (see, however,~\cite{Bambeck:2005bz}). Considering that the \n\ mass eigenstates have a fundamental character, and that they allow a consistent quantum field theory treatment in the Standard Model (with a minimal extension to account for neutrino mass terms),  we infer that it should be possible to apply the steps of the Hawking radiation derivation outlined above to the \n\ mass eigenstates, independently of the Hawking radiation of other particle fields. Therefore, it might be natural to expect the \n\ to be emitted as mass eigenstates. However, considering the lack of in-depth addresses of the question, to be conservative here we  \emph{assume} that BHs emit neutrinos in mass eigenstates. 

This hypothesis has interesting immediate implications. First, black holes might be the only \emph{emitters} of \n\ mass eigenstates \footnote{Other \ns\ (e.g., the cosmological relic neutrinos and the supernova relic \n\ background) might be found as mass eigenstates in the universe today, however they are the result of the progressive decoherence of the wavepackets of \ns\ that were originally produced as flavor eigenstates. See, e.g. \cite{Hannestad:2010kz}.}. After being produced, neutrinos from BHs propagate as free particles, either in the curved spacetime close to the BH or in the Minkowski spacetime far from it.  Since they are eigenstates of the vacuum Hamiltonian (refraction effects are negligible in the Early universe, in the absence of exotic physics~\cite{Diaz:2015aua}), the \n\ mass states do not oscillate. They interact via the weak interaction through their mixing with the flavor eigenstates. In particular, if a neutrino in a mass eigenstate $\nu_a$ reaches a detector, the probability to detect it by producing a charged lepton $\ell_\alpha$ will be proportional to $|U_{\alpha a}|^2$, see Eq. \eqref{eq:CCLag}.  

\begin{figure}[t]
\centering
\includegraphics[width=0.8\textwidth]{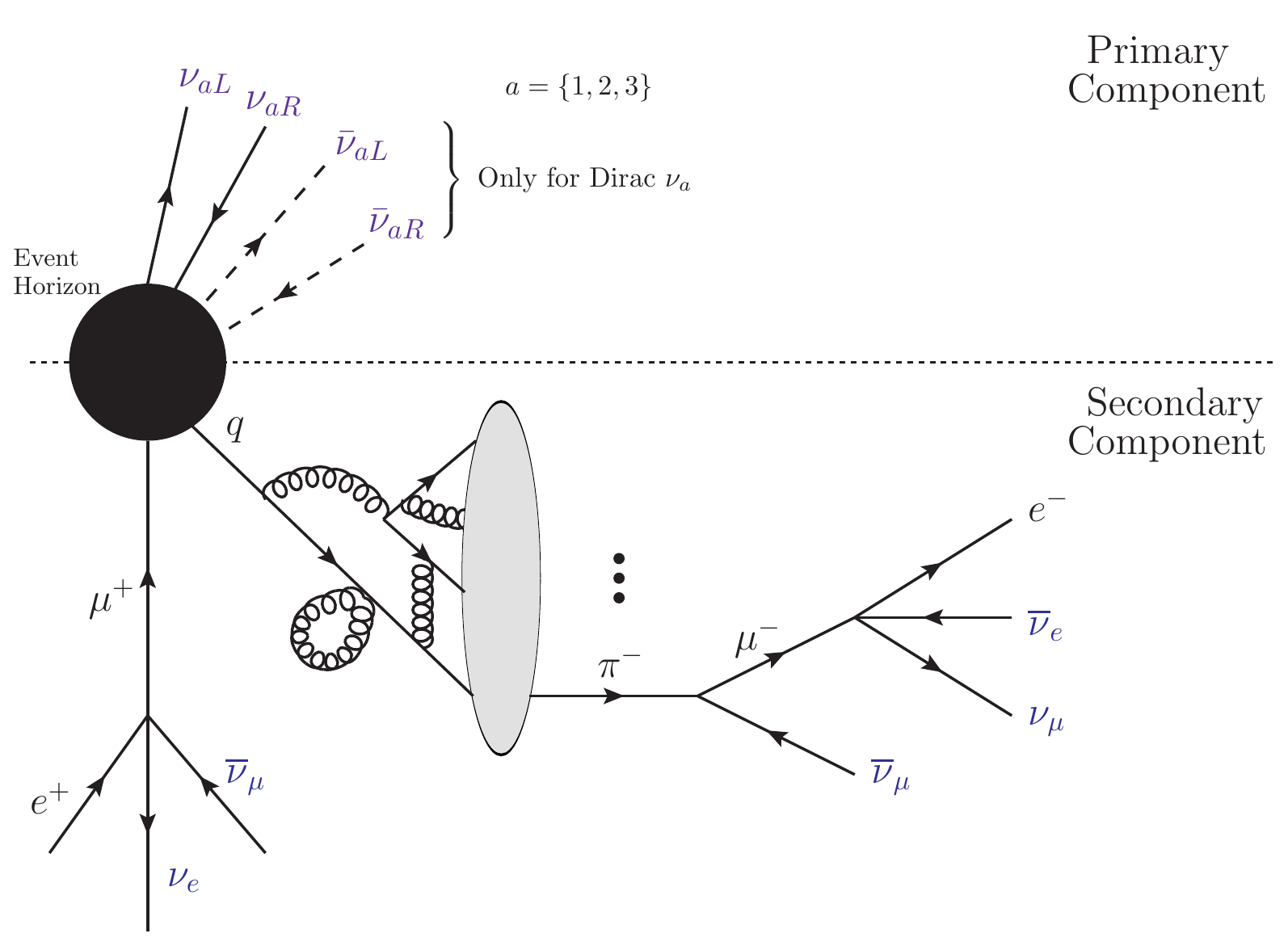}
\caption{An illustration of the two different types of neutrino emission, primary and secondary, from a PBH.}
\label{fig:cart}
\end{figure}

Aside from whether the emitted states are mass or flavor eigenstates, neutrino production from PBH evaporation will depend on the \n\ unknown fermionic nature (Dirac or Majorana). This because the Hawking spectrum depends on the internal degrees of freedom of the emitted particle, see illustration in fig. \ref{fig:cart}\footnote{Note that a nonzero initial abundance of RH Majorana neutrinos produced from the evaporation could lead to generation of matter-antimatter asymmetry, a PBH driven Leptogenesis, see \cite{Fujita:2014hha,Hamada:2016jnq,Morrison:2018xla}.}.  Moreover, due to the absence of helicity suppression, both LH and RH neutrinos and their antineutrinos are produced with equal rates in the Dirac case \cite{Unruh:1976fm}.

\begin{figure}[t]
\centering
\includegraphics[width=0.55\textwidth]{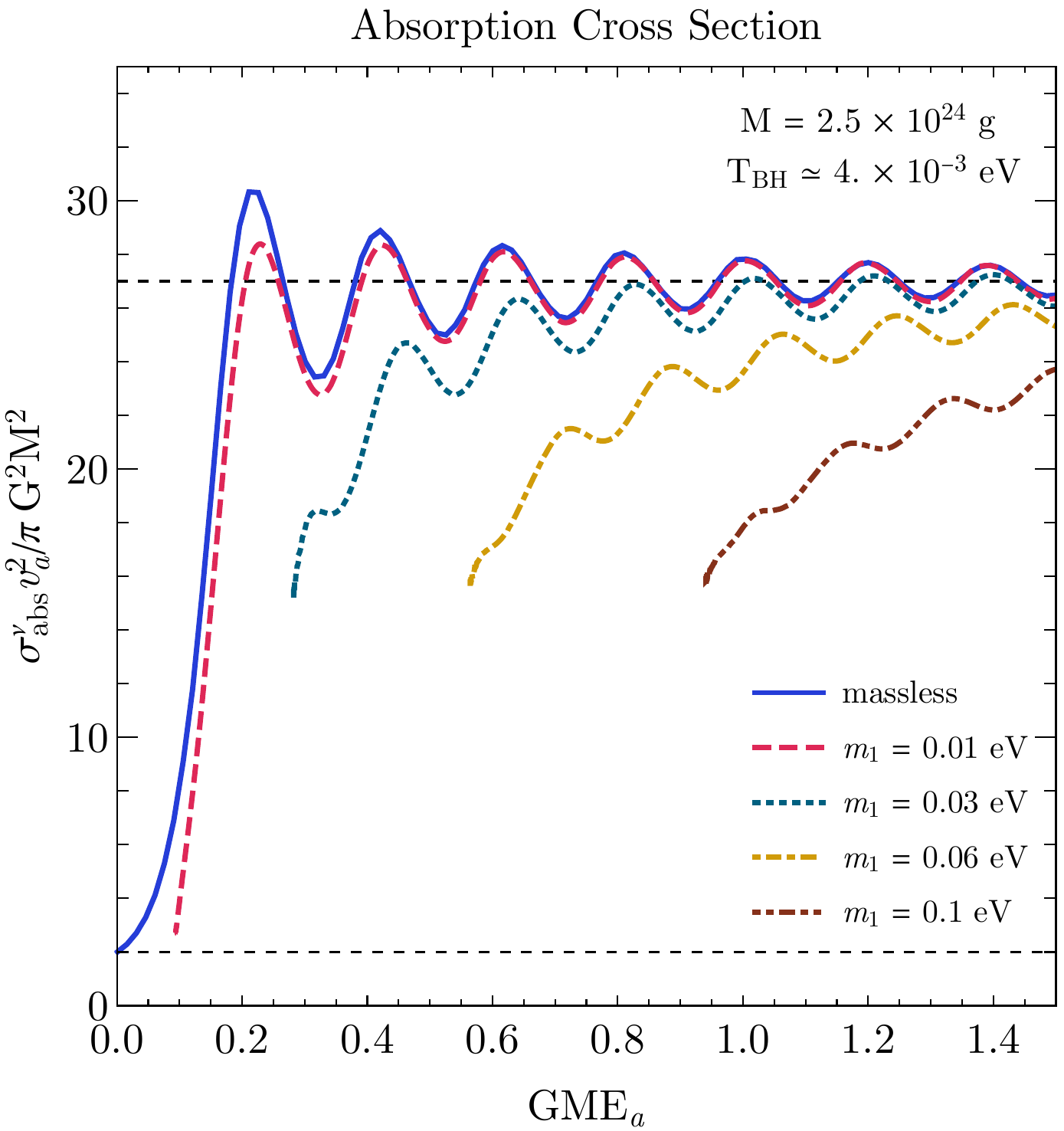}
\caption{Absorption cross section times velocity squared of the lightest neutrino for a BH with mass $M=2.5\times 10^{24}$ g, as function of the neutrino energy (in units of $G^{-1}M^{-1}$) for $0\eV \leq m_1 \leq 0.01\eV$. The horizontal dashed lines indicate the results (for massless neutrinos) in the limits $G M E_a\ll 1$ 
($\sigma^\nu_{\rm abs}\simeq 2\pi G^2M^2$) and $G M E_a\gg 1$ ($\sigma^\nu_{\rm abs}\simeq 27\pi G^2M^2$), see text.}
\label{fig:absxsec}
\end{figure}
%

To fix the ideas, let us consider the \n\ mass spectrum with normal mass ordering and the lightest neutrino mass set to be $m_0=0.01$ eV. Using parameters from recent fits to \n\ oscillation data \cite{Esteban:2018azc}, the three masses, $m_a$, are then
$m_1=m_0$, $m_2 \approx 1.32\times 10^{-2}$ eV, $m_3 \approx 5.12\times 10^{-2}$ eV.
The emission rate of \ns\ with momentum between $p$ and $p+dp$ by a Schwarzschild BH is \cite{Hawking:1974sw, Hawking:1974rv, Page:1976df, MacGibbon:1990zk, Halzen:1995hu}\footnote{Here a factor $dE_a/dp = p/E_a$ was used to obtain $dN/dp dt$ from the (more familiar) expression of $dN/dE_a dt$.}
\begin{align}\label{eq:nuHS}
	\frac{d^2 N_{\nu}}{dp\,dt}=\sum_{a=1,2,3}\frac{g_a^N}{2\pi^2}\frac{\sigma^\nu_{\rm abs}(M,p,m_a)\,}{\exp[E_a(p)/T_{\rm BH}]+1}\frac{p^3}{E_a(p)},
\end{align}
with $E_a$ being the total energy of a neutrino $\nu_a$ and $g_a^N$ ($N$=Dirac or Majorana) the number of internal degrees of freedom. The quantity $\sigma_{\rm abs}^\nu$ is the cross section for the absorption of a state $\nu_a$ with momentum $p$ by the BH, dependent on the gravitational coupling, $\alpha_g^a = G M m_a$. It presents a oscillatory behavior due to the contribution of the different partial waves \cite{Doran:2005vm}. In the limit $G M E_a \ll 1$, $\sigma^\nu_{\rm abs}$ is dominated by the first partial wave, approaching a value of $\sigma^\nu_{\rm abs} \to 2\pi G^2 M^2$ in the case of massless neutrinos \cite{Unruh:1976fm,Page:1976df}. For  $G M E_a\gg 1$, the absorption cross section tends to the geometric optics limit, $\sigma^\nu_{\rm abs} \to 27\pi^2G^2M^2$. In Fig.\ \ref{fig:absxsec}, we show the absorption cross-section times neutrino velocity squared, $\sigma^\nu_{\rm abs} v_a^2$, as function of
$GME_a$, for representative values of the lightest neutrino mass assuming a BH with $M=2.5\times 10^{24}{\rm\ g}$. Notice that the curves start at values corresponding to the zero momentum limit ($G M E_a=\alpha_g^a$) of $\sigma^\nu_{\rm abs} v_a^2/\pi G^2M^2$ \cite{Page:1977um}.
%
\begin{figure}[t]
\centering
\includegraphics[width=0.925\textwidth]{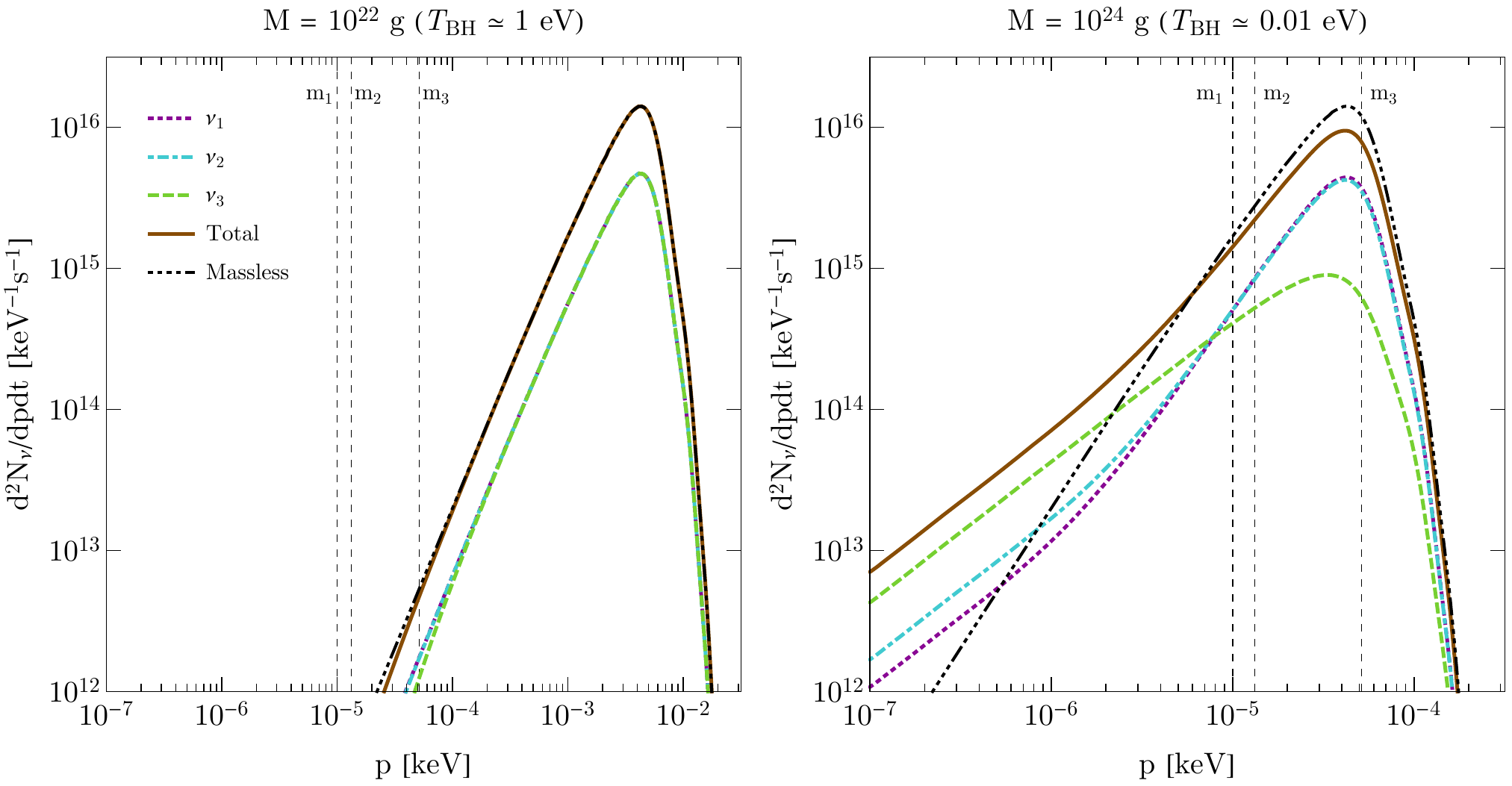}
\caption{Primary Hawking emission spectrum as function of the \n\ momentum for massless neutrinos and Majorana \ns\  with masses $m_1=10^{-2}$ eV, $m_2 \simeq 1.3\times 10^{-2}$ eV, $m_3 \simeq 5\times 10^{-2}$ eV and PBHs with $M=10^{22}$ g (left) and $M=10^{24}$ g (right).  We also show the contribution of each mass eigenstate.}
\label{fig:HEspectrumex}
\end{figure}
%

In Fig.\  \ref{fig:HEspectrumex} we show examples of massless and Majorana mass eigenstate \n\ emission rate (in terms of the neutrino momentum, $p$) for values $M=10^{22}\ {\rm g}\ (T_{\rm BH} \approx 1\ {\rm eV} -$ left) and $ M= ~10^{24}\ {\rm g}\ (T_{\rm BH}\approx 0.01\ {\rm eV} -$ right). As expected, in the first case Majorana and massless \n\ spectra fully coincide. For the higher mass, the spectra of the mass eigenstates show modifications compared to the massless case, due to the mass effects in the Hawking spectrum \cite{MacGibbon:1990zk}. Also, the emission rate becomes exponentially suppressed when $E_a\gtrsim T$, as expected from a Fermi-Dirac distribution.

Finally, we can summarize the effect of neutrino masses and the fermionic nature in the primary neutrino emission from a Schwarzschild BH as follows:
\begin{itemize}
    \item \emph{Majorana neutrinos}: neutrino mass effects are important for the emission when $m_a\sim T_{\rm BH}$, and there are two degrees of freedom per mass eigenstate, LH and RH neutrinos, that can be emitted, similar to the massless case.
    \item \emph{Dirac neutrinos}: neutrino mass effects are important for $m_a\sim T_{\rm BH}$, and there are four possible degrees of freedom per mass eigenstate that can be emitted, LH and RH neutrinos and antineutrinos. 
\end{itemize}

\subsection{Effects on the PBH evaporation}
\label{sub:eveff}

\begin{figure}[t]
\centering
\includegraphics[width=\textwidth]{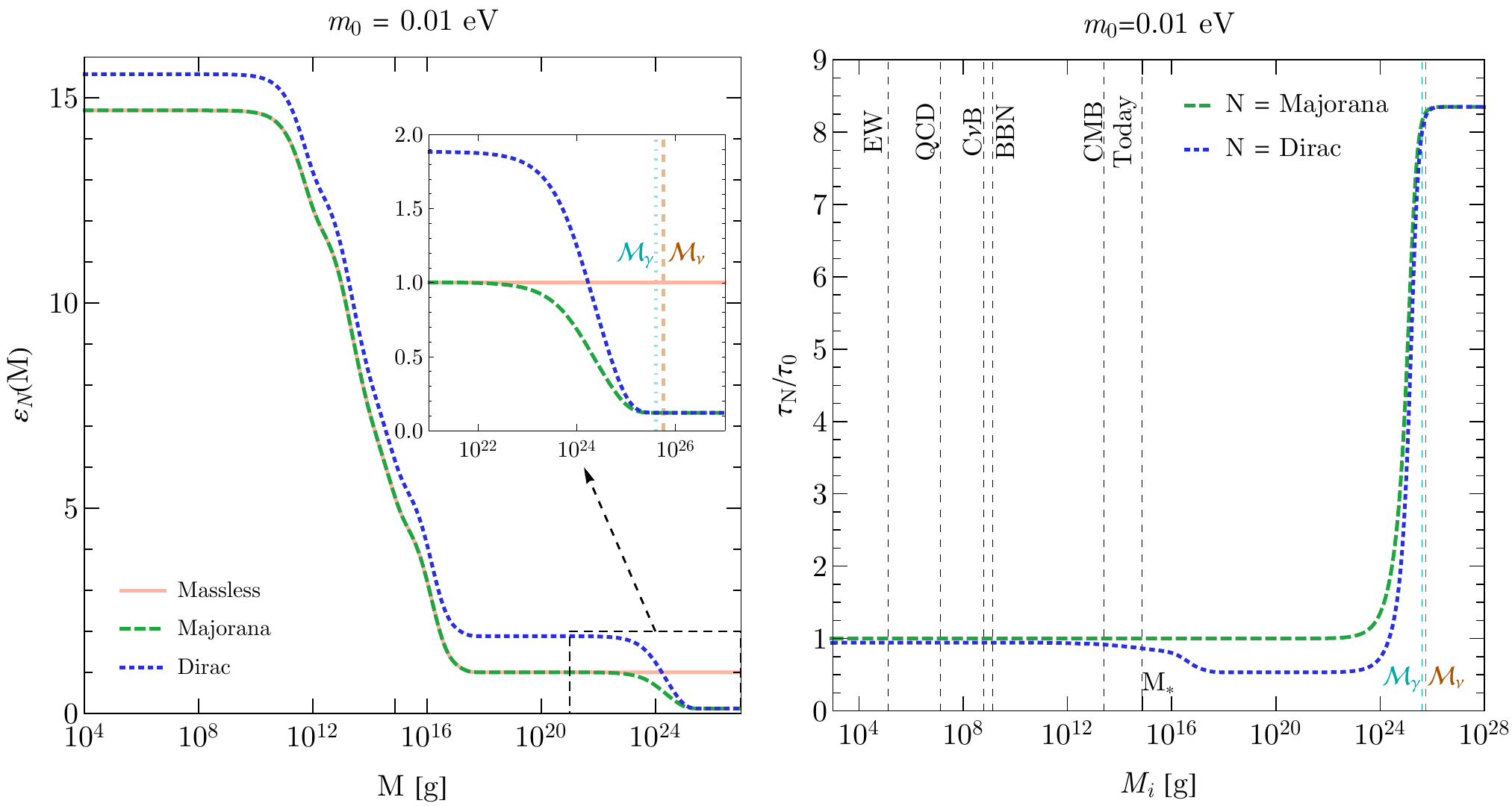}
\caption{The evaporation function $\varepsilon_N(M)$ (left panel, with zoom-in inset) and the ratio of black hole lifetimes for massive and massless \ns\ (right panel), in the cases of Majorana and Dirac neutrinos. The evaporation function in the massless case is also presented in the left panel. The dashed lines in the right panel indicate the PBH mass for which the lifetime would be equal to the different epochs of the early Universe (assuming the standard cosmological model): Electroweak phase transition (EW), QCD phase transition, the neutrino decoupling ($\CNB$), the beginning of the Big Bang Nucleosynthesis (BBN), and the age of the Universe, denoted by $M_*$. In both panels,  ${\cal M}_\gamma$(${\cal M}_\nu$) indicate the PBH mass values for which the black hole has the same temperature as the CMB ($\CNB$).}
\label{fig:fM}
\end{figure}

Let us now discuss how the primary emission of massive \ns\ affects the time evolution of a black hole. 
Due to evaporation, a PBH loses mass with a rate given by~\cite{MacGibbon:1990zk,MacGibbon:1991tj,Carr:2009jm} 
\begin{align}\label{eq:MEq}
	\dot{M}&=-\sum_{j}\frac{g_j}{2\pi^2}\int_0^\infty \frac{\sigma^{s_j}_{\rm abs}(M,p)\,p^2}{\exp[E_j(p)/T_{\rm BH}]-(-1)^{2s_j}}\,p\,dp\notag\\
	&= -5.34\times 10^{25} ~{\rm g\, s^{-1}}\,\varepsilon_N(M)\left(\frac{\rm 1\ g}{M}\right)^{2} ~,
\end{align}
where the sum is done over all the particle species $j$ ($j=l,q,a,g,W,Z,H$, corresponding to the SM set of particles: leptons, quarks, neutrinos mass eigenstates, gluons, W, Z and the Higgs boson). For each species $j$, the quantities $E_j$, $g_j$, and $s_j$ indicate the energy, internal degrees of freedom  and spin; $\sigma^{s_j}_{\rm abs}(M,p)$ are the spin- and momentum-dependent absorption cross sections (see Sec. \ref{sub:emission}). In \eqref{eq:MEq}, $\varepsilon_N(M)$ is the evaporation function, which 
can be expressed as \cite{MacGibbon:1991tj}:
\begin{align}\label{eq:fMnu}
	\varepsilon_N(M) &= 2f_1+4f_{1/2}^{ 1}\left\{\sum_{\ell=e,\mu,\tau}\exp\left[-\frac{M}{\beta_{1/2} M_\ell}\right] +3\sum_{q}\exp\left[-\frac{M}{\beta_{1/2} M_q}\right]\right\}
	\notag\\
	&\quad +2\,\eta_\nu^N f_{1/2}^{ 0}\sum_{a=1,2,3}\exp\left[-\frac{M}{\beta_{1/2} M_a}\right]\notag\\
	&\quad +16f_1\exp\left[-\frac{M}{\beta_1 M_g}\right]	\notag\\
	&\quad +3f_1\left\{2\exp\left[-\frac{M}{\beta_1 M_W}\right]+\exp\left[-\frac{M}{\beta_1 M_Z}\right]\right\}+f_0\exp\left[-\frac{M}{\beta_0 M_H}\right],
\end{align}
where $M_j$ is  the mass of a PBH with a temperature equal to the mass of the particle $j$
\begin{align*}
	M_j= \frac{1}{8\pi G m_j}\approx \left(\frac{\rm 1.06\ GeV}{m_j}\right)\cdot 10^{13}\ {\rm g}~,
\end{align*}
and $\eta_\nu^N$ accounts for the difference between Dirac and Majorana degrees of freedom,
\begin{align*}
	\eta_\nu^N = \begin{cases}
		2 & \text{for } N = \text{Dirac}\\
		1 & \text{for } N = \text{Majorana}
	\end{cases}.
\end{align*}
The factors $f_{0,1,2}$, $f_{1/2}^{ (0,1)}$ appearing in \eqref{eq:MEq} describe the contribution to $\varepsilon_N(M)$ per degree of freedom depending on the spin and the charge of the emitted particle. The spin-dependent parameters $\beta_{0,1/2,1,2}$ are fixed such that the emitted power of a black hole with $M=\beta_s M_j$ is maximum at $p = m_j$ \cite{MacGibbon:1990zk,MacGibbon:1991tj}
\begin{align*}
	\beta_s = \begin{cases}
		2.66 & \text{for } s = 0\\
		4.53 & \text{for } s = \frac{1}{2}\\
		6.04 & \text{for } s = 1
	\end{cases}, \quad
	f_s = \begin{cases}
		0.267 & \text{for } s = 0\\
		0.060 & \text{for } s = 1\\
		0.007 & \text{for } s = 2
	\end{cases}, \quad
	f_{1/2}^{q} = \begin{cases}
		0.147 & \text{for } q = 0\ {\rm (neutral)}\\
		0.142 & \text{for } q = 1\ {\rm (charged)}
	\end{cases}.
\end{align*}
Note that here  $\varepsilon(M)$ is defined so that $\varepsilon=1$ for massless \ns\ and in the high $M$ limit ($M\gtrsim 10^{17}$ g, i.e., only neutrinos and photons emitted, see fig. \ref{fig:fM}). 

%
By integrating the mass loss rate, eq.\ \eqref{eq:MEq}, we obtain the lifetime $\tau_{N}$ of a PBH of initial mass $M_i$  \cite{MacGibbon:1991tj,Carr:2009jm}. To illustrate its dependence on the mass and nature of the \n, in fig. \ref{fig:fM} we show the ratio $\tau_{N} (M_i)/\tau_0(M_i)$ (with $\tau_0$ being the lifetime in the massless \n\ case), and the function $\varepsilon_N(M)$ for massless, Dirac and Majorana \ns.
In the figure, we observe three different regimes, depending on the initial PBH mass (or, equivalently, the initial temperature, $T_{\rm BH}^{i}$):

\begin{itemize}

\item The {\it low mass regime}. for $M_i\lesssim 10^{16}$ g ($T_{\rm BH}^{i} \gtrsim 1$ MeV), the neutrino emission is always accompanied by the emission of other SM particles. The emitted neutrinos are relativistic, so results for Majorana and massless neutrinos coincide. In the Dirac case, the additional degrees of freedoms enhance the evaporation function up to $10\%$, resulting in a comparable  ($\sim 10\%$ or less) shortening of the lifetime compared to the massless/Majorana case. 

\item The {\it intermediate mass regime}.  for $10^{16}$ g $\lesssim M_i \lesssim 10^{24}$ g ($  10^{-3}~{\rm eV} \lesssim T_{\rm BH}^{i} \lesssim 1$ MeV), a PBH  only radiates neutrinos and photons for most of its life. The \ns\ are mostly relativistic, implying only minor differences between the Majorana and massless cases.  For Dirac \ns, the extra degrees of freedom increase the initial emissivity by up to a factor of 2, relative to the massless case, with a corresponding reduction of the lifetime by almost half.

\item The {\it high mass regime}.  if $M_i \gtrsim 10^{25}$ ($T_{\rm BH}^{i} \lesssim 10^{-3}$ eV) a PBH evolution is dominated by photons, with strong mass-suppression of the \n\ emission. Due to the photon-domination, the evaporation function (and therefore the lifetime)  is approximately the same for Dirac and Majorana \ns.  As $M_i$ increases, the lifetime ratio starts to converge to a value of $8.35$. Let us notice, however, that PBHs with masses
\begin{subequations}
\begin{align}
    {\cal M}_\gamma &\equiv \frac{1}{8 \pi G T_{\rm CMB}} \approx 4\times 10^{25}\ {\rm g}\\
    {\cal M}_\nu &\equiv \frac{1}{8 \pi G T_{\CNB}} \approx 5.65\times 10^{25}\ {\rm g},
\end{align}
\end{subequations}
would have a temperature equal to the temperature of the Cosmic Microwave Background (CMB) or the Cosmic Neutrino Background ($\CNB$), respectively. Thus, any PBH with mass $M_i > {\cal M}_\gamma~({\cal M}_\nu)$ absorbs more photons (neutrinos) from the CMB ($\CNB$) than it emits. Therefore, as $M_i$ approaches ${\cal M}_\gamma$, the PBH evolution is no longer described by evaporation only, as photon absorption from the CMB starts to dominate.
\end{itemize}

An important question is how the effect of the \n\ mass and nature on the PBH evolution could affect cosmology. To address it, in fig. \ref{fig:fM} we show the values of $M_i$ for which the PBH lifetime would be equal to the beginning of the different cosmological epochs (assuming the standard cosmological model, see figure caption). The  PBH mass corresponding to a lifetime equal to the age of the Universe is found to be $M_i=M_*\approx\{7.5, 7.5, 7.8\}\times 10^{14}$ g, for massless, Majorana and Dirac neutrinos, respectively. It falls in the low $M_i$ regime, where mass effects are negligible.  Therefore, we conclude that the neutrino mass effect are significant only for PBHs which are still present in the Universe today.

\newpage

\section{PBH evaporation in the early Universe: the case of Dirac neutrinos }\label{sec:const}

\subsection{Constraints on the initial PBH fraction}
\label{sub:const}

Dirac neutrinos can be introduced in the SM framework with the minimal addition of singlet right-handed states, $\nu_{aR}$. The Yukawa interaction terms, $\mathscr{L}_Y=-Y_\nu^{ab} \overline{L_L^a}\widetilde{H}\,\nu_{b R}$, generate neutrino masses of $\sim{\cal O}({\rm eV})$ after the Electroweak symmetry breaking if the Yukawa couplings $Y_\nu^{ab}$ are of order $\sim{\cal O}(10^{-12})$. Thus, RH states are not produced thermally in the Early Universe in the minimal scenario\footnote{Furthermore, note that an initial non-thermal RH neutrino density cannot thermalize with the primordial plasma \cite{Chen:2015dka}.} \cite{Antonelli:1981eg,Chen:2015dka}. Nevertheless, PBH evaporation could emit an important population of RH neutrinos, modifying the evolution of the Universe. This could impose a limit on the initial PBH fraction since the effective number of neutrino species, $N_{\rm eff}$, has been constrained to be $N_{\rm eff} = 2.99\pm 0.17$ ($\Delta N_{\rm eff}\equiv N_{\rm eff} - N_{\rm eff}^{\rm SM} <0.28$ at 2$\sigma$ C.\ L.) by CMB + BAO measurements \cite{Aghanim:2018eyx}. Furthermore, future experiments that intend to measure $N_{\rm eff}$ with higher precision could improve the constraints \cite{Abazajian:2019oqj}. These are the South Pole Telescope SPT-3G ($\Delta N_{\rm eff}<0.12$ at 2$\sigma$ C.\ L.) \cite{Benson:2014qhw}, the CMS Simmons Observatory ($\Delta N_{\rm eff}<0.05-0.07$ at 1$\sigma$ C.\ L.) \cite{Abitbol:2019nhf} and the CMB Stage-4 (CMB-S4) experiments ($\Delta N_{\rm eff}<0.06$ at 95\% C.\ L.) \cite{Abazajian:2019eic}.

\begin{figure}[t]
\centering
\includegraphics[width=0.65\textwidth]{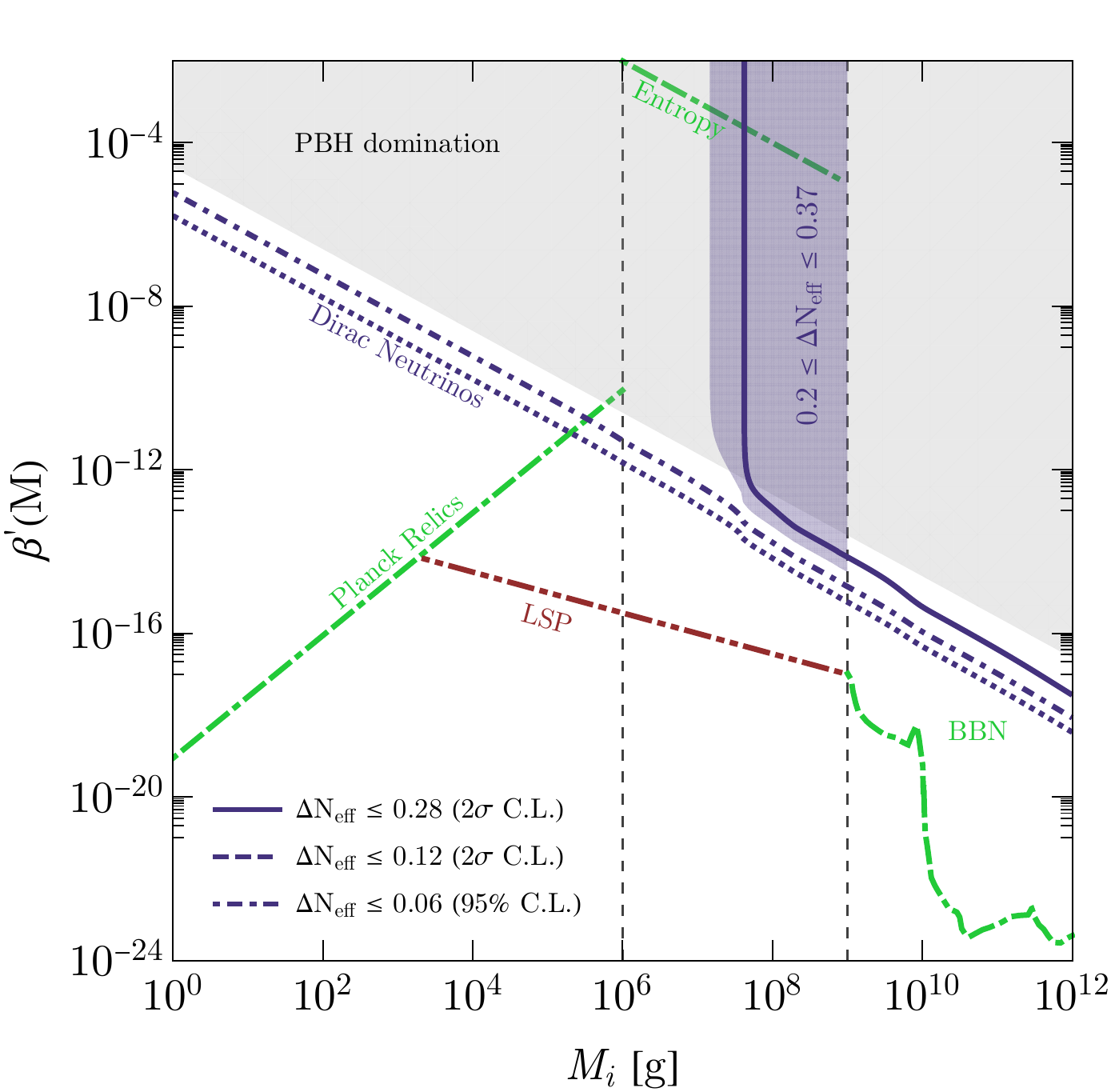}
\caption{Constraints on the initial PBH fraction $\beta^\prime$ as a function of the initial PBH mass, $M_i$, due to the emission of RH states in the case of Dirac neutrinos. We consider limits stemming from constraints on $\Delta N_{\rm eff}$ (see legend): (i) the current limit of $\Delta N_{\rm eff} \leq 0.28$, and the expected sensitivities of future experiments, specifically, (ii) the South Pole Telescope/CMS Simmons Observatory, $\Delta N_{\rm eff} \leq 0.12$ and (iii) the CMB Stage-4, $\Delta N_{\rm eff} \leq 0.06$ (see text). The purple shaded region corresponds to the PBH parameters that produce $0.2 \lesssim \Delta N_{\rm eff} \lesssim 0.37$, values that can ease the tension on the Hubble measurements, see \protect\cite{Riess:2016jrr,Aghanim:2018eyx,Escudero:2019gzq,Hooper:2019gtx}.  Bounds from entropy generation, the production of a 100 GeV Lightest Supersymmetric Particle (LSP), Planck relics and BBN have been taken from \protect\cite{Carr:2009jm} (legends on curves). The gray shaded region indicates the parameters which lead to PBH dominated era in the Early Universe, see Eq.\ \eqref{eq:domination}.}
\label{fig:Limbeta}
\end{figure}

Considering an initially radiation-dominated Universe, let us assume that a PBH po\-pu\-la\-tion was formed with a monochromatic PBH mass distribution, and that the initial PBH mass is dependent on the particle horizon mass as $M_i=4\pi\gamma\rho^i_{\rm tot} H^{-3}/3$, with $\rho^i_{\rm tot}$ the total energy density, $H$ the Hubble parameter and $\gamma = (3\sqrt{3})^{-1}$, a dimensionless parameter related to the gravitational collapse \cite{Carr:1975qj,Carr:2009jm}\footnote{We have checked that our results are only mildly dependent on the gravitational collapse factor value.}.  The temperature of the Universe at the time in which PBHs form, $T_{\rm f}$, is then
\begin{align}
    T_{\rm f} = \left(\frac{45}{16\pi^3 G^3}\right)^\frac{1}{4}\,g_{*}(T_{\rm f})^{-\frac{1}{4}}\gamma^{\frac{1}{2}}M_i^{-\frac{1}{2}},
\end{align}
with $g_{*}(T_f)$ the number of relativistic degrees of freedom at the PBH formation time.  We parametrize the initial PBH density fraction $\rho_{\rm PBH}^i$ to the total energy as \cite{Carr:2009jm}
\begin{align}
    \beta^\prime=\gamma^{\frac{1}{2}}\left(\frac{g_*(T_{\rm f})}{106.75}\right)^{-\frac{1}{4}}\frac{\rho_{\rm PBH}^i}{\rho_{\rm tot}^i}. 
\end{align}
In Fig.\  \ref{fig:Limbeta} we summarize upper limits on $\beta^\prime$ (taken from \cite{Carr:2009jm}) for massless \ns\ and for the mass region $1\ {\rm g}\lesssim M_i \lesssim 10^{12}$ g. Constraints are strong for masses $M_i\gtrsim 10^{9}$ g, since the final stages of the evaporation would occur during the BBN \cite{Carr:2009jm}. For $10^6\ {\rm g}\lesssim M_i\lesssim 10^{9}$ g, a model dependent bound has been obtained considering the production of the lightest superpartner (LSP) in a Supersymmetric scenario \cite{Green:1999yh}. In the same region there exists a model independent but weaker constraint corresponding to the modification of the photon-to-baryon ratio by additional photons from the evaporation \cite{Zeldovich:1977aa}. For lower PBH masses, $M_i\lesssim 10^6$ g, the possible production of Planck-mass relics introduces another constraint \cite{MacGibbon:1987my,Barrow:1992hq,Carr:1994ar,Nozari:2005ah,Chen:2004ft,Barrau:2003xp,Chen:2002tu,Alexeyev:2002tg,Carr:2009jm}. Nevertheless, such bound relies on the assumption that BHs do not evaporate completely, and it can introduce additional complications \cite{Susskind:1995da}. Although any consideration of the BH evolution when its mass gets closer to the Planck mass is certainly precarious, here we assume that PBHs evaporate completely. 

To consistently model the production of RH neutrinos by PBHs, and its impact on the Universe evolution, we consider the set of Friedmann equations for the energy densities of PBH ($\rho_{\rm PBH}$), SM radiation ($\rho_{\rm R}$) and RH neutrinos ($\rho_{\nu_{\rm R}}$) \cite{Barrow:1991dn,Hooper:2019gtx,Lennon:2017tqq} 
\begin{subequations}\label{eq:FEqs}
\begin{align}
    \dot{\rho}_{\rm PBH} + 3 H \rho_{\rm PBH} &= \frac{\dot{M}}{M}\rho_{\rm PBH},\\
    \dot{\rho}_{\rm R} + 4 H \rho_{\rm R} &= -\frac{\varepsilon_{\rm SM}(M)}{\varepsilon_{\rm D}(M)}\frac{\dot{M}}{M}\rho_{\rm PBH},\\
    \dot{\rho}_{\rm \nu_R} + 4 H \rho_{\rm \nu_R} &=  -\frac{\varepsilon_{\nu_{\rm R}}}{\varepsilon_{\rm D}(M)}\frac{\dot{M}}{M}\rho_{\rm PBH},\\
    H^2&=\frac{8\pi G}{3}(\rho_{\rm PBH}+\rho_{\rm R}+\rho_{\rm \nu_R})~,
\end{align}
\end{subequations}
with the standard definition $H=\dot{a}_t/a_t$, $a_t=a(t)$ the scale factor at the time $t$, $\varepsilon_{SM}(M)$ the evaporation function for the SM degrees of freedom only, and $\varepsilon_{\nu_{\rm R}} = 6 f_{1/2}^0$, the contribution of the 6 (2 per mass eigenstate) neutral additional states in the case of Dirac neutrinos. Note that the system of equations \eqref{eq:TEq} is fully general. It accounts for the possibility that, depending on the initial PBH fraction, the PBHs dominate the energy density before the final stages of their evaporation, changing the evolution of the Universe and leading to a non-standard cosmology \cite{Hooper:2019gtx}. 

Since PBH evaporation changes the radiation energy density, the entropy is no longer conserved. Therefore, to describe the evolution of the temperature in the Universe, $T_{\rm U}$, we use the evolution of the entropy density \cite{Bernal:2019lpc,Arias:2019uol}
\begin{align}
    \dot{s}_{\rm R} + 3Hs_{\rm R} = -\frac{\varepsilon_{\rm SM}(M)}{\varepsilon_{\rm D}(M)}\frac{\rho_{\rm PBH}}{T_{\rm U}}\frac{\dot{M}}{M}~, 
\end{align}
which gives an evolution equation for the temperature,
\begin{align}\label{eq:TEq}
    \frac{\dot{T}_{\rm U}}{T_{\rm U}} &= -\frac{1}{\Delta}\left\{ H + \frac{\varepsilon_{\rm SM}(M)}{\varepsilon_{\rm D}(M)}\frac{\dot{M}}{M} \frac{g_{*}(T_{\rm U})}{g_{*S}(T_{\rm U})} \frac{\rho_{\rm PBH}}{4(\rho_{\rm R} + \rho_{\nu_{\rm R}})}\right\}~.
\end{align}
Here the $\Delta$ parameter describes the dependence of the entropic relativistic degrees of freedom on the temperature \cite{Bernal:2019lpc,Arias:2019uol}:
\begin{align}
    \Delta = 1 + \frac{T_{\rm U}}{3 g_{*S}(T_{\rm U})}\frac{dg_{*S}(T_{\rm U})}{dT_{\rm U}}~.
\end{align}
Thus, we have to solve the full system of the Friedmann equations eqs.\ \eqref{eq:FEqs}, together with the temperature evolution eq.\ \eqref{eq:TEq} and the mass lose rate, eq.\ \eqref{eq:MEq} from the formation time $t_f$, until the evaporation time, approximately equal to the PBH lifetime, $t_{\rm ev}\approx\tau_D$. We obtain the temperature of the plasma at which the PBH disappearance occurs, denoted as evaporation temperature $T_{\rm EV}$, by evaluating eq.\ \eqref{eq:TEq} at the evaporation.

After complete evanescence of the PBH,  we want to quantify the modification of the effective number of neutrino species at the matter-radiation equality. To do so, we relate the SM radiation and RH neutrino energy densities between the evaporation and the matter-radiation equality considering their dependence on the scale factors at such epochs, $a_{\rm EV}$, $a_{\rm EQ}$, respectively \cite{Hooper:2019gtx},
\begin{subequations}
\begin{align}
    \frac{\rho_{\nu_{\rm R}}(T_{\rm EQ})}{\rho_{\nu_{\rm R}}(T_{\rm EV})} &= \left(\frac{a_{\rm EV}}{a_{\rm EQ}}\right)^4, \\ \frac{\rho_{\rm R}(T_{\rm EQ})}{\rho_{\rm R}(T_{\rm EV})} &=
    \frac{g_{*}(T_{\rm EQ})T_{\rm EQ}^4}{g_{*}(T_{\rm EV})T_{\rm EV}^4}
    = \left(\frac{g_{*}(T_{\rm EQ})}{g_{*}(T_{\rm EV})}\right)\left(\frac{a_{\rm EV}}{a_{\rm EQ}}\right)^4\left(\frac{g_{*S}(T_{\rm EV})}{g_{*S}(T_{\rm EQ})}\right)^\frac{4}{3}.
\end{align}
\end{subequations}
with $T_{\rm EQ}\approx 0.75$ eV. Here the factors of $g_*, g_{*S}$ account for the possible reheating of the thermal bath due to particle decays. Now, using the definition of $\Delta N_{\rm eff}$
\begin{align*}
	\Delta N_{\rm eff}&=\frac{\rho_{\nu_{\rm R}}(T_{\rm EQ})}{\rho_{\nu_{\rm L}}(T_{\rm EQ})}~,
\end{align*}
with $\rho_{\nu_{\rm L}}(T_{\rm EQ})$ the active neutrino energy density, we have \cite{Hooper:2019gtx}
\begin{align}\label{eq:Neffg}
    \Delta N_{\rm eff} = \left\{\frac{8}{7}\left(\frac{4}{11}\right)^{-\frac{4}{3}}+N_{\rm eff}^{\rm SM}\right\} 
    \frac{\rho_{\rm \nu_R}(T_{\rm EV})}{\rho_{\rm R}(T_{\rm EV})}
    \left(\frac{g_*(T_{\rm EV})}{g_*(T_{\rm EQ})}\right)
    \left(\frac{g_{*S}(T_{\rm EQ})}{g_{*S}(T_{\rm EV})}\right)^{\frac{4}{3}},
\end{align}
with $N_{\rm eff}^{\rm SM} = 3.045$ the effective number of relativistic species in the SM \cite{deSalas:2016ztq}. Note that the expression in eq.\ \eqref{eq:Neffg} is valid for any value of the initial PBH fraction, as the solutions of the Friedmann equations (eqs. \eqref{eq:FEqs}) are directly dependent on the initial condition on $\beta^\prime$. In the scenario in which there was a PBH-dominated era, the entire population of SM particles plus RH neutrinos would come from the evaporation \cite{Hooper:2019gtx}. A PBH-domination would occur if the initial fraction is larger than \cite{Hooper:2019gtx}
\begin{align} 
\label{eq:domination}
    \beta^\prime&\gtrsim 2.5\times 10^{-14} \left(\frac{g_*(T_{\rm f})}{106.75}\right)^{-\frac{1}{4}} \left(\frac{M_i}{10^8\ {\rm g}}\right)^{-1}\left(\frac{\varepsilon_{D}(M_i)}{15.35}\right)^{\frac{1}{2}}~;
\end{align}
the corresponding region in the $M_i$-$\beta^\prime$ plane is shown in Fig. \ref{fig:Limbeta} (gray shading). 

Let us first discuss the change in $N_{\rm eff}$ due to the three RH \ns\ for the PBH-dominated case, Eq. (\ref{eq:domination}).In this limit, $\Delta N_{\rm eff}$ depends only on $M_i$ through the evaporation temperature, $T_{EV}$, as follows:
\begin{align}
\label{eq:Dneffappr}
    \Delta N_{\rm eff} \approx 0.772\, \frac{g_*(T_{\rm EV})}{g_{*S}(T_{\rm EV})^\frac{4}{3}}\,,
\end{align}
where we have used that the RH neutrino and radiation energy densities are directly related to their respective contributions to the evaporation function~\cite{Hooper:2019gtx}. From Eq. (\ref{eq:Dneffappr}), we find that $\Delta N_{\rm eff}$ increases with $M_i$, from a minimum value of $\Delta N_{\rm eff}\simeq 0.14$, which is realized  for PBHs that completely evaporated before the EW phase transition ($T_U \gtrsim T_{EW}\sim 10^2$ GeV), where the effect of RH \ns\ to the relativistic degrees of freedom is diluted by the contribution of all the other particles. 
Increasing $M_i$ corresponds to longer lived PBHs, which survive until epochs of lower temperature, where the contribution of the RH \ns\ to $N_{\rm eff}$ is progressively more important. 
For $M_i \gtrsim 4.3\times 10^7\g$, we have $\Delta N_{\rm eff} \gtrsim  0.28$, which is in tension with the PLANCK constraint at $2\sigma$ C.L. or more, thus resulting in a constraint on $\beta^\prime$ in this mass range (solid line in Fig. \ref{fig:Limbeta}). In this region of the parameter space, PBH evaporation occurred after the QCD phase transition ($T_U \lesssim T_{QCD}\sim 0.1$ GeV).  Our results for the PBH-dominated regime are in agreement with those in \cite{Hooper:2019gtx}. 
%
%

For values of $\beta^\prime$ outside the PBH-domination regime, $\Delta N_{\rm eff}$ can be calculated from Eq. (\ref{eq:Neffg}), and depends on both $M_i$ and $\beta^\prime$, in a way that the constraint  on $\beta^\prime$ due to the PLANCK limit becomes stronger with increasing $M_i$. A broad summary of the situation is given in Fig.\  \ref{fig:Limbeta}. The figure shows that 
the constraint on $\beta^\prime$ is improved by $\sim 10$ orders of magnitude (compared to the Entropy limit) for the region $4.3\times 10^7\ {\rm g} \lesssim M_i \lesssim 10^9$ g (BBN constraints dominate for $M_i \gtrsim 10^9$ g), where the largest allowed value of $\beta^\prime$ is  $10^{-14} \lesssim \beta^\prime_{max} \lesssim 10^{-12}$. Note that our area of exclusion  extends into the region where PBHs did not dominate the evolution of the Universe. In Fig. \ref{fig:Limbeta}, we also illustrate possible stronger constraints that may derive from future, more stringent limits on $\Delta N_{\rm eff}$, in the assumption that \ns\ are Dirac fermions. Such future constraints could reach  initial masses as low as $M_i\sim 1$ g and fractions of $\beta^\prime \gtrsim 10^{-6}$.

\subsection{PBHs and indications of excess radiation}
\label{sub:excess}

Until now, we have discussed experimental results on $N_{\rm  eff}$ from the perspective of restricting the allowed PBH parameter space. But what would be the implications if an excess in  $N_{\rm  eff}$ (i.e., $\Delta N_{\rm eff}>0$) is established? In that case, an attractive explanation could be found in PBHs, under the sole, minimal assumption that \ns\ be Dirac fermions.  

As an illustration, let us consider the recent claims that extra radiation, at the level of $0.2 \lesssim \Delta N_{\rm eff} \lesssim 0.5$, can alleviate the tension between measurements of the Hubble parameter at early and late times \cite{Riess:2016jrr,Aghanim:2018eyx,Escudero:2019gzq,Hooper:2019gtx,Vagnozzi:2019ezj}. In our specific scenario in which RH neutrinos are produced from PBH evaporation, a contribution up to $\Delta N_{\rm eff} \sim 0.37$ can be generated while satisfying all the other constraints on $\beta^\prime$.  Specifically (see fig. \ref{fig:Limbeta}, purple shaded region),  $0.2 \lesssim \Delta N_{\rm eff} \lesssim 0.37$ is possible for PBH masses in the range $10^7\lesssim M_i \lesssim 10^9$ g and $\beta^\prime \gtrsim 10^{-13}$. 

Interestingly, in the region of the parameters where PBH evaporation with Dirac \ns\ contributes significantly to $\Delta N_{\rm eff}$, models involving PBHs and new (non-neutrino) light degrees of freedom would be restricted (compared to the case of massless \ns). This, however, has some caveats: if new models assume Majorana neutrinos or Dirac neutrinos with other interactions, the limits derived here would be different.

\section{Diffuse neutrino flux from PBHs}\label{sec:diffflux}

An interesting question is whether the diffuse flux of neutrinos from PBHs at Earth is detectable. To answer, let us discuss how the radiated \ns\ evolve to the present time. 
When PBHs are first formed, the average energy of their emitted \ns\ is (see Eq. \eqref{eq:TBH}) $   \langle E_\nu \rangle \sim T_{\rm BH} \sim {\cal O}(10^{21})\eV\left({1\g}/{M_i}\right)$. Thus, for PBHs with $M_i\gtrsim 10^{21}\g$, neutrinos are emitted as non-relativistic or semi-relativistic fermions of LH and RH helicity. Because helicity is conserved in their  propagation, the \ns\ remain helicity eigenstates at all times. At their arrival at Earth, all the \n\ states would then have a left-chiral component, and they would interact weakly, allowing for a possible detection.  

For $M_i\lesssim 10^{21}\g$, neutrinos are emitted as ultrarelativistic particles, for which helicity and chirality coincide. The RH neutrino states propagate by free streaming, and -- due to helicity conservation -- arrive at Earth as RH helicity eigenstates, suffering only redshift of energy. If they are non-relativistic at arrival (for $M_i\lesssim 10^6\g$),  their non-zero left-chiral component will make them detectable via the weak interaction. A similar fate applies to the \ns\ produced as LH helicity eigenstates, provided that they are always decoupled from the plasma (i.e., $M_i\gtrsim 10^9\g$, corresponding to emission after \n\ decoupling, see Fig. \ref{fig:fM}).  If that is not the case ($M_i\lesssim 10^9\g$), then the emitted LH \ns\ would equilibrate with the C$\nu$B, and be effectively lost to detection. 

Previous works on detecting \ns\ from PBHs \cite{Carr:1976zz, Halzen:1995hu, Bugaev:2000bz,  Bugaev:2002yt, Dave:2019epr} have considered the active neutrino flux with energies $E_\nu\gtrsim 1$ MeV, and have included both primary and secondary emissions. 
Here, we will consider only the regimes where secondary emission is absent or suppressed, so primary emission dominates. This is the case for: 

\begin{itemize}
    \item  the flux of Dirac (LH + RH) neutrinos and antineutrinos or Majorana (LH + RH) neutrinos from PBHs that are still present in the Universe  today ($M_i>M_\ast \simeq 7.5~10^{14}$ g). These PBHs would contribute to a fraction of the dark matter (DM), which  is subject to several constraints (see e.g., \cite{Carr:2009jm,Carr:2017jsz,Katz:2018zrn}). However, since our purpose is to consider the possible observational effects of nonzero neutrino masses in the evaporation and possible constraints on $\beta^\prime$ from neutrino measurements, we will assume that all DM is constituted by PBHs.
    
    \item the RH Dirac neutrino flux for PBH that have already completely evaporated ($M_i< M_\ast$, including the regime $M_i<10^9~{\rm g}$). For this scenario we will assume that the Universe had a PBH-dominated era at some point, corresponding to the region of the parameters space in Eq. (\ref{eq:domination}).  
\end{itemize}

We compute the flux by integrating the Hawking spectrum of a PBH with initial mass $M_i$ over the time $t$, including redshift effects, as follows \cite{Kim:1999iv,Bugaev:2000bz}: 
\begin{align}
	\frac{d\Phi_{\rm PBH}^\nu}{dp_0}
	&=\int_{t_i}^{\min(t_0,\tau)}  dt\, \frac{d\Omega}{4\pi}\, \frac{a_0}{a_t}\left(\frac{a_i}{a_0}\right)^3\, \frac{\rho_{\rm PBH}^i}{M_i}\, \frac{d^2 N_\nu}{dp\, dt}(M(t),p_0 \,a_0/ a_t),
\end{align}
with $a_0,\ a_i$ the scale factor at the present and at the PBH formation time, respectively; $p_0$ is the neutrino momentum today, redshifted from the initial momentum $p$. The integration is performed between $t_i$ (the formation time) and over the black hole lifetime (that is, until the time $t_i+\tau \simeq \tau$), or until the present time, $t_0$, if the PBH has not completely evaporated yet. The ratio $({a_i}/{a_0})$ (and, analogously, $a_0/ a_t$) can be found using the equation:
\begin{align}
    \frac{a_i}{a_0}&=\left(\frac{a_i}{a_{\rm EV}}\right)\left(\frac{a_{\rm EV}}{a_{\rm EQ}}\right)\left(\frac{a_{\rm EQ}}{a_0}\right)\notag\\
    &= \left(\frac{a_i}{a_{\rm EV}}\right)\left(\frac{g_{*S}(T_{\rm EV})}{g_{*S}(T_{\rm EQ})}\right)^\frac{1}{3}\left(\frac{T_{\rm EV}}{T_{\rm EQ}}\right)\left(1+z_{\rm EQ}\right),
\end{align}
with $z_{\rm EQ}$ being the redshift corresponding to $a_{\rm EQ}$. To obtain the ratio $a_i/a_{\rm EV}$, we use the solutions of the Friedmann equations (eqs. \eqref{eq:FEqs}).

\begin{figure}[t]
\centering
\includegraphics[width=\textwidth]{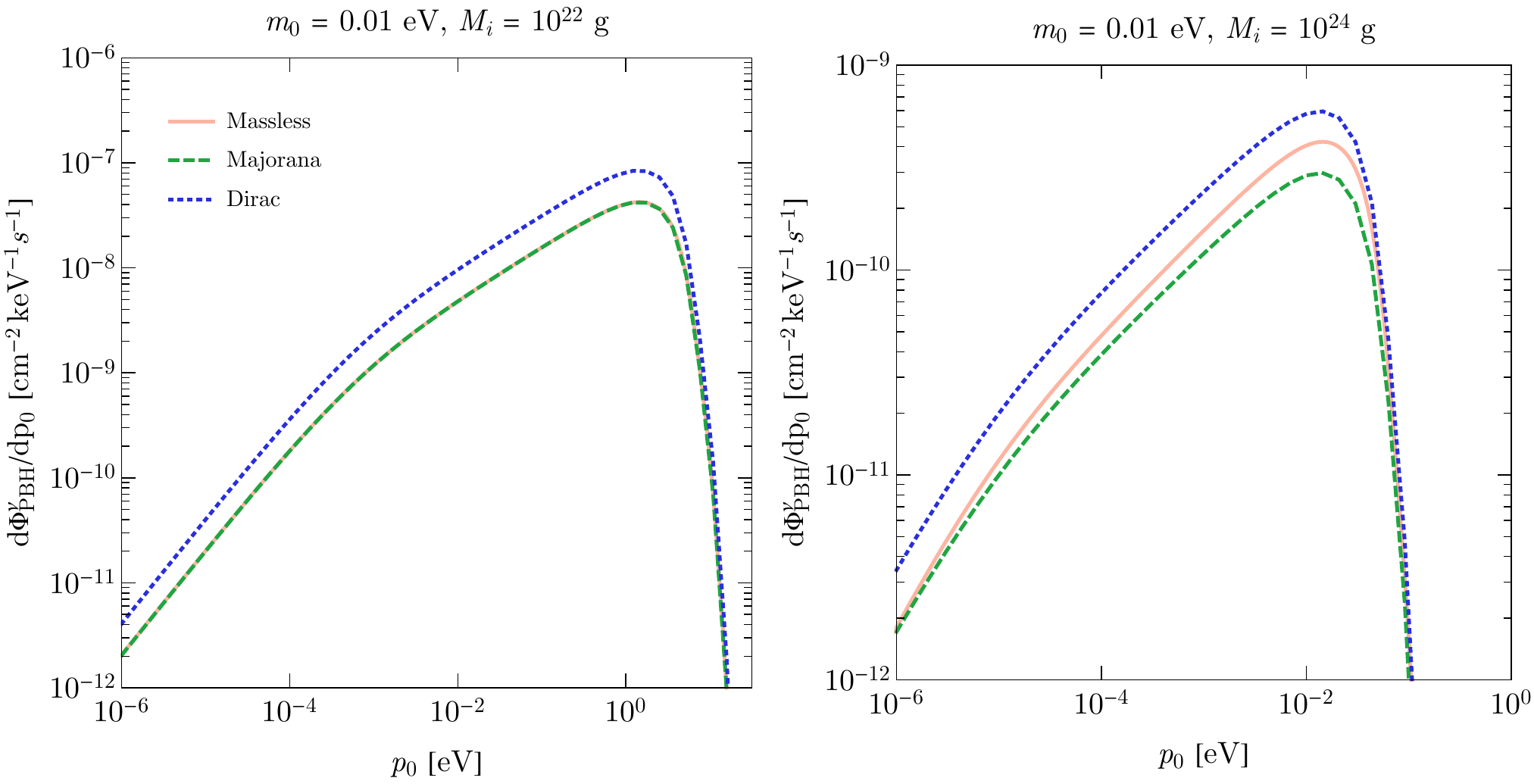}
\caption{Spectrum of the PBH neutrino flux (considering all DM as made of PBH) for a PBH mass of $M_i=10^{22}$ g (left) and $M_i=10^{24}$ g (right).}
\label{fig:PBHsFMD}
\end{figure}

\begin{figure}[t]
\centering
\includegraphics[width=\textwidth]{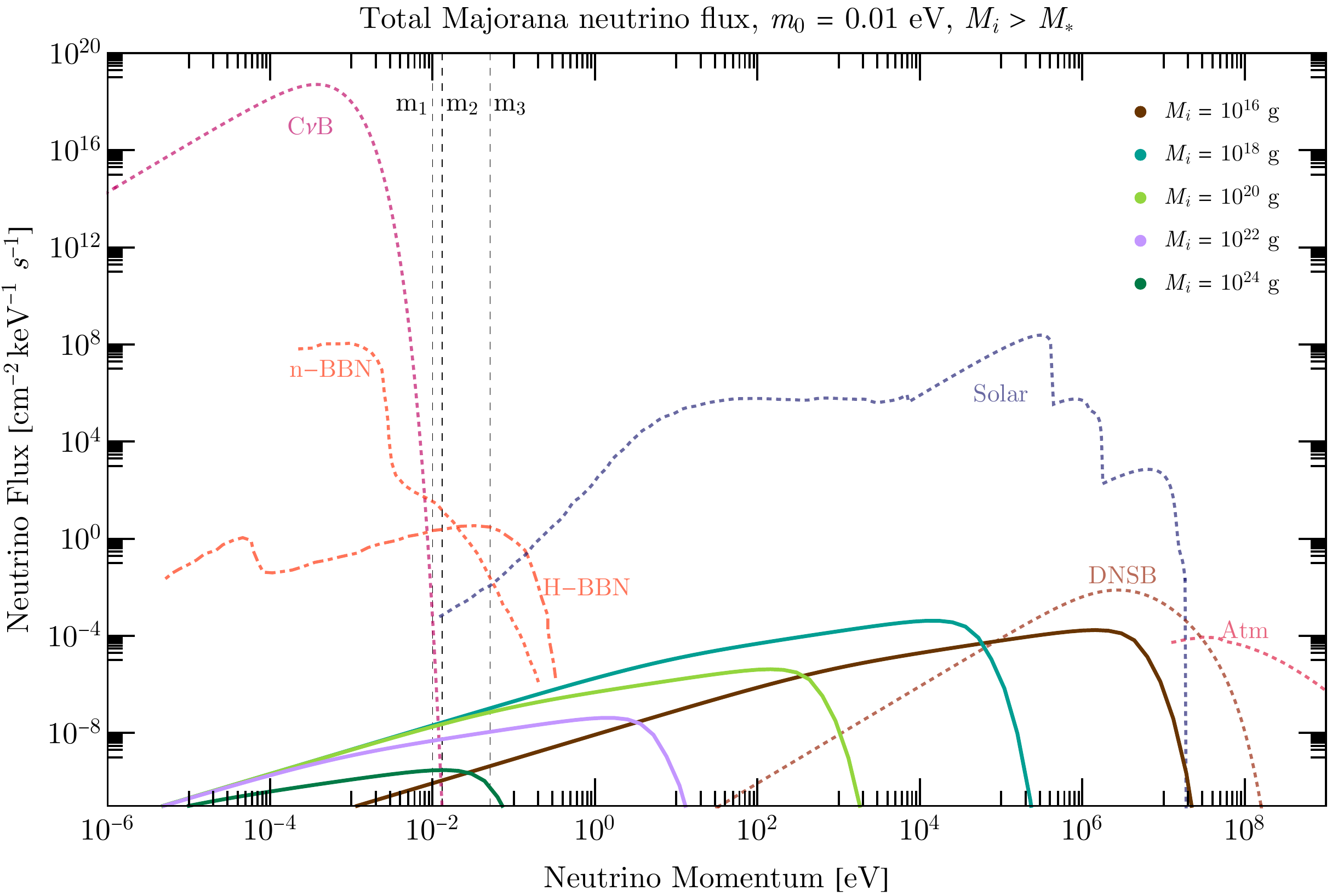}
\caption{Diffuse Majorana neutrino flux for values of $\beta^\prime$ that saturate current limits (see Fig.\  \ref{fig:Limbeta}) for PBHs with $M_i > M_*$. For comparison, we also present the solar neutrino flux \cite{Bahcall:2000nu,Bahcall:2005va,Vitagliano:2017odj}, low-energy atmospheric flux \cite{Battistoni:2005pd,Honda:2011nf}, the diffuse supernova neutrino background (DSNB) \cite{Beacom:2010kk}, the $\CNB$ flux \cite{Dolgov:2002wy,Dolgov:2008hz,Lesgourgues:2006nd,Quigg:2008ab}, and neutrinos from decays of neutron and tritium produced at the BBN \cite{Ivanchik:2018fxy}. The dashed vertical lines indicate the assumed neutrino masses.} 
\label{fig:PBHsSA}
\end{figure}

As a first step, we would like to understand if a measurement of the diffuse fluxes could shed some light on the neutrino nature. In Fig. \ref{fig:PBHsFMD} we show the spectrum of the total neutrino diffuse flux for PBHs in the intermediate mass regime (see Sec. \ref{sub:eveff}): $M_i=10^{22}$ g and $M_i=10^{24}$ g. As expected, the fluxes in the Dirac and Majorana scenarios differ by a factor of 2. For the case $M_i=10^{24}$ g, neutrino mass effects in both absorption cross section and Hawking spectrum introduce differences with respect to the massless case, see Fig. \ref{fig:HEspectrumex}. Thus, the diffuse fluxes contain the information of the neutrino mass and nature. Nevertheless, other neutrino fluxes would constitute a background to searches of neutrinos from PBHs. When we compare the Majorana neutrino diffuse flux  for PBHs that still exist today with fluxes from other sources (see fig. \ref{fig:PBHsSA}), we find that the latter dominate, making a possible detection difficult in this case.

%

%
\begin{figure}[t]
\centering
\includegraphics[width=\textwidth]{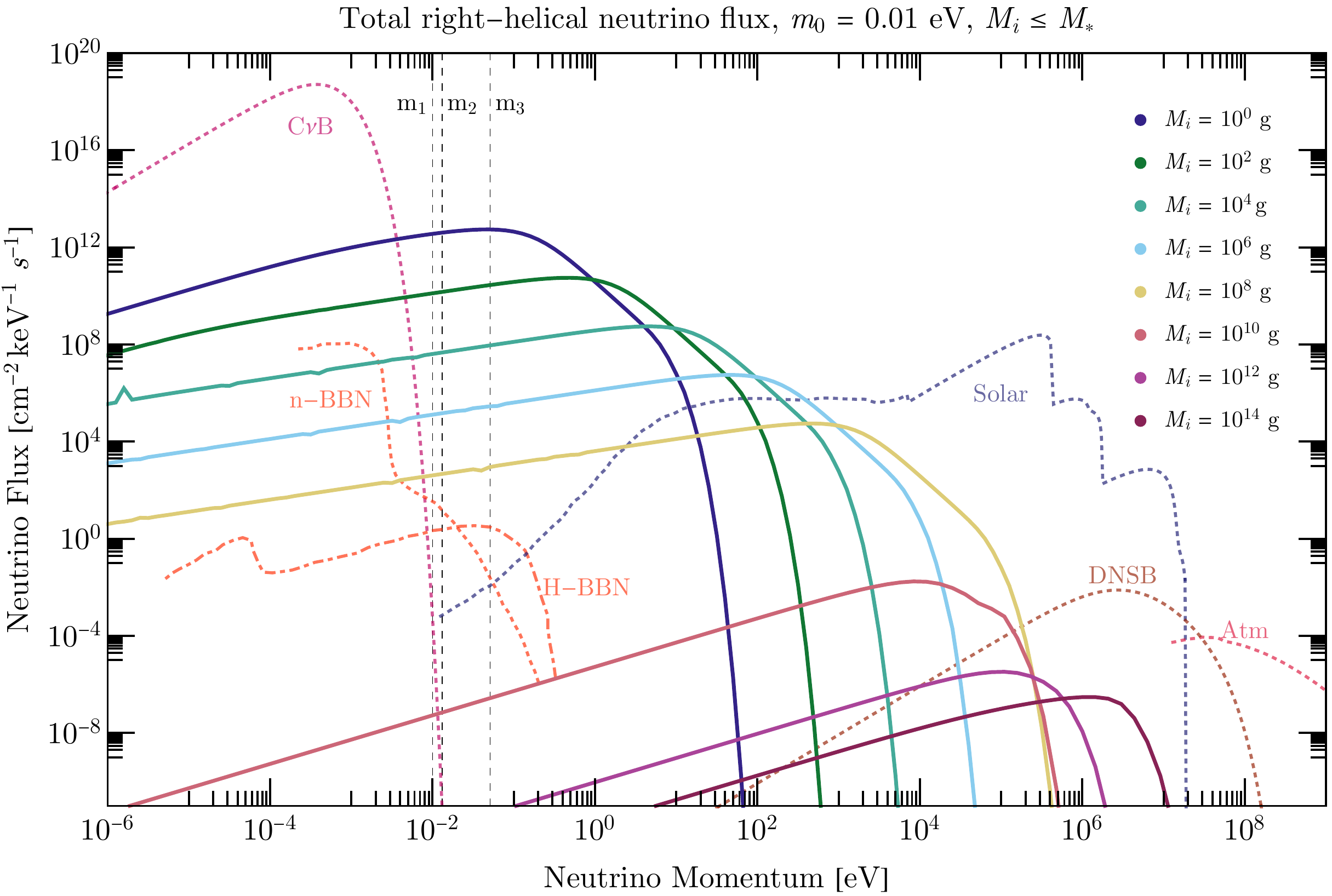}
\caption{Same as Fig.\  \ref{fig:PBHsSA}, but for right-helical neutrinos and PBH with masses $M_i\le M_*$.}
\label{fig:PBHsSB}
\end{figure}

Coming now to the case $M_i\le M_*$, results for the RH \n\ flux are shown in Fig. \ref{fig:PBHsSB}.  For $10^{10}$ g $\lesssim M_i \lesssim$ $10^{14}$ g, we find that such flux is suppressed due to the bounds on the PBH initial fraction from BBN and gamma ray fluxes, see Fig.\ \ref{fig:Limbeta}. For masses $M_i \lesssim$ $10^{8}$ g, the RH neutrino fluxes are comparable to, or even exceed, the fluxes from other sources for $3\times 10^{-3}\eV \lesssim p_0\lesssim 1\keV$. Therefore, we identify  this region of the parameter space as the most promising for detection. In this context, let us consider the detectability at a realistic facility. 
PTO\-LE\-MY \cite{Betti:2019ouf} is a proposed experiment with the capability to detect non-relativistic \ns\ (with the cosmic neutrino background being the main candidate, see also \cite{Long:2014zva,McKeen:2018xyz,Chacko:2018uke}) via capture on tritium, $\nu_a +\,^3{\rm H} \to\,^3{\rm He} +e ^-$ \cite{Weinberg:1962zza,Cocco:2007za,Cocco:2009rh}. 

Following refs.\ \cite{Cocco:2007za,Cocco:2009rh,Long:2014zva}, we have estimated the capture rate for \ns\ from PBHs to be $\Gamma^\nu_{\rm PBH} \sim 10^{-2}\ [{\rm kg - year}]^{-1}$, for  $M_i=1$ g (the most optimistic case shown in Fig. \ref{fig:PBHsSB}). Considering that PTOLEMY will operate with 0.1 kg of tritium,  detection appears impractical for the time being. Still, it may be worth to explore other possible detection mechanisms beyond the simplest ones. This is left for future work. 

\section{Conclusions}\label{sec:conc}

After the discovery of gravitational waves, the astrophysics of PBHs has seen a renewed interest in the literature. In this work, we have studied the phenomenology of neutrinos from PBHs, in the light of the recent advances in neutrino physics. The neutrino emission from BH is completely different from the familiar weak interaction production, given that the Hawking process is a purely quantum effect in a gravitational background. We have adopted a scenario where the neutrinos are not emitted as flavor eigenstates, but as states with definite masses, and discussed how the primary neutrino emission is different in the case of Dirac and Majorana fermions because the particle emission from PBHs depends on the internal degrees of freedom.

We have obtained the result that the PBH lifetime depends on the neutrino fermionic nature and mass, in such a way that for PBHs masses of $10^{18}\ {\rm g}\lesssim M_i\lesssim 10 ^{24}$ g, the  lifetime in the Dirac case is half the one for Majorana neutrinos. For larger masses, the lifetimes becomes $\sim 8.35$ times the value in the case of massless neutrinos. However, such dramatic effect can not be tested directly, because black holes in the mass region $M_i \gtrsim 10^{15}~{\rm g}$ have not completely evaporated yet, and thus are still present in the Universe. For masses smaller than $10^{12}$ g, the difference in the PBH lifetime between Dirac and Majorana is reduced to be $\sim$10 \%, because the other SM particles are emitted, and the relative neutrino contribution is small.

If neutrinos are Dirac particles, a significant non-thermal population of RH neutrinos can be present at the BBN or the CMB production epoch, so that $N_{\rm eff}$ can be larger than its standard value. Assuming a monochromatic PBH mass distribution where the initial mass is related to the particle horizon mass at the formation, we have shown that the current cosmological bound on $\Delta N_{\rm eff}$ implies a constraint on the initial fraction of PBHs for initial black hole masses $4.3\times 10^7\g\lesssim M_i \lesssim 10^{9}\g$. Future experiments could improve this constraint, and extend it to black hole masses as low as $\sim 1$ g. 

We have identified an interesting region of the parameter space -- masses $10^7\lesssim M_i\lesssim 10^5$ g, and  initial fraction of $\beta^\prime \gtrsim 10^{-13}$ -- where $\Delta N_{\rm eff}$ could be large enough to ease the tension between early and late measurements of the Hubble constant: $0.2\lesssim \Delta N_{\rm eff}\lesssim  0.37$.
This last result is a minimal realization of a recently discussed scenario where PBHs emit light, sterile particles \cite{Hooper:2019gtx}. 

Taking into account all the existing constraints, we have estimated the largest possible diffuse flux of RH \ns\ from PBHs at Earth. We found that the most promising scenario for detectability is for black holes with mass $M_i \sim {\mathcal O}(1)\g$. If the PBHs dominated the evolution of the Universe, they could cause a flux of non-relativistic right-helical \ns\ that exceeds all other \n\ fluxes in the momentum window $p_0\sim 3\times 10^{-3}~ - ~10^3$ eV.  These \ns\ have a non-zero left chiral component, so in principle they are detectable.  Considering detection via absorption on tritium, as in the proposed PTOLEMY experiment, we find that for the most optimistic PBH parameters, a detection rate of one event per decade would require 1 kg of tritium, which is currently unrealistic. Moreover, it may be difficult to distinguish a signal due to PBHs from the one due to the $\CNB$. Nevertheless, we think that investigating experimentally achievable methods of detection of RH neutrinos from PBHs will be an interesting direction to pursue.

\acknowledgments

We thank Gordan Krnjaic and Maulik Parikh for very helpful discussions. YFPG would like to thank for the kind hospitality received at the Department of Physics of Arizona State University where part of this work was completed. CL acknowledges funding from the National Science Foundation grant number PHY-1613708. Fermilab is operated by the Fermi Research Alliance, LLC under contract No. DE-AC02-07CH11359 with the United States Department of Energy. 

\appendix

\bibliographystyle{JHEP}
\bibliography{references}{}

\providecommand{\href}[2]{#2}\begingroup\raggedright\begin{thebibliography}{100}

\bibitem{Abbott:2016blz}
{\bf LIGO Scientific, Virgo} Collaboration, B.~P. Abbott et~al., {\it
  {Observation of Gravitational Waves from a Binary Black Hole Merger}},  {\em
  Phys. Rev. Lett.} {\bf 116} (2016), no.~6 061102,
  [\href{http://arxiv.org/abs/1602.03837}{{\tt arXiv:1602.03837}}].

\bibitem{Hawking:1971ei}
S.~Hawking, {\it {Gravitationally collapsed objects of very low mass}},  {\em
  Mon. Not. Roy. Astron. Soc.} {\bf 152} (1971) 75.

\bibitem{Carr:1974nx}
B.~J. Carr and S.~W. Hawking, {\it {Black holes in the early Universe}},  {\em
  Mon. Not. Roy. Astron. Soc.} {\bf 168} (1974) 399--415.

\bibitem{Carr:2005zd}
B.~J. Carr, {\it {Primordial black holes: Do they exist and are they useful?}},
   in {\em {59th Yamada Conference on Inflating Horizon of Particle
  Astrophysics and Cosmology Tokyo, Japan, June 20-24, 2005}}, 2005.
\newblock \href{http://arxiv.org/abs/astro-ph/0511743}{{\tt astro-ph/0511743}}.

\bibitem{Khlopov:2008qy}
M.~{\relax Yu}. Khlopov, {\it {Primordial Black Holes}},  {\em Res. Astron.
  Astrophys.} {\bf 10} (2010) 495--528,
  [\href{http://arxiv.org/abs/0801.0116}{{\tt arXiv:0801.0116}}].

\bibitem{Carr:2009jm}
B.~J. Carr, K.~Kohri, Y.~Sendouda, and J.~Yokoyama, {\it {New cosmological
  constraints on primordial black holes}},  {\em Phys. Rev.} {\bf D81} (2010)
  104019, [\href{http://arxiv.org/abs/0912.5297}{{\tt arXiv:0912.5297}}].

\bibitem{Ali-Haimoud:2016mbv}
{Ali-Ha\"{\i}moud, Yacine and Kamionkowski, Marc}, {\it {Cosmic microwave
  background limits on accreting primordial black holes}},  {\em Phys. Rev.}
  {\bf D95} (2017), no.~4 043534, [\href{http://arxiv.org/abs/1612.05644}{{\tt
  arXiv:1612.05644}}].

\bibitem{Bird:2016dcv}
{Bird, Simeon, Cholis, Ilias, Muñoz, Julian B., Ali-Ha\"{\i}moud, Yacine,
  Kamionkowski, Marc, Kovetz, Ely D., Raccanelli, Alvise and Riess, Adam G.},
  {\it {Did LIGO detect dark matter?}},  {\em Phys. Rev. Lett.} {\bf 116}
  (2016), no.~20 201301, [\href{http://arxiv.org/abs/1603.00464}{{\tt
  arXiv:1603.00464}}].

\bibitem{Carr:2016drx}
B.~Carr, F.~Kuhnel, and M.~Sandstad, {\it {Primordial Black Holes as Dark
  Matter}},  {\em Phys. Rev.} {\bf D94} (2016), no.~8 083504,
  [\href{http://arxiv.org/abs/1607.06077}{{\tt arXiv:1607.06077}}].

\bibitem{Inomata:2017okj}
K.~Inomata, M.~Kawasaki, K.~Mukaida, Y.~Tada, and T.~T. Yanagida, {\it
  {Inflationary Primordial Black Holes as All Dark Matter}},  {\em Phys. Rev.}
  {\bf D96} (2017), no.~4 043504, [\href{http://arxiv.org/abs/1701.02544}{{\tt
  arXiv:1701.02544}}].

\bibitem{Sasaki:2018dmp}
M.~Sasaki, T.~Suyama, T.~Tanaka, and S.~Yokoyama, {\it {Primordial black
  holes—perspectives in gravitational wave astronomy}},  {\em Class. Quant.
  Grav.} {\bf 35} (2018), no.~6 063001,
  [\href{http://arxiv.org/abs/1801.05235}{{\tt arXiv:1801.05235}}].

\bibitem{Hawking:1974rv}
S.~W. Hawking, {\it {Black hole explosions}},  {\em Nature} {\bf 248} (1974)
  30--31.

\bibitem{Hawking:1974sw}
S.~W. Hawking, {\it {Particle Creation by Black Holes}},  {\em Commun. Math.
  Phys.} {\bf 43} (1975) 199--220. [,167(1975)].

\bibitem{Carr:2017jsz}
B.~Carr, M.~Raidal, T.~Tenkanen, V.~Vaskonen, and H.~Veermäe, {\it {Primordial
  black hole constraints for extended mass functions}},  {\em Phys. Rev.} {\bf
  D96} (2017), no.~2 023514, [\href{http://arxiv.org/abs/1705.05567}{{\tt
  arXiv:1705.05567}}].

\bibitem{Lennon:2017tqq}
O.~Lennon, J.~March-Russell, R.~Petrossian-Byrne, and H.~Tillim, {\it {Black
  Hole Genesis of Dark Matter}},  {\em JCAP} {\bf 1804} (2018), no.~04 009,
  [\href{http://arxiv.org/abs/1712.07664}{{\tt arXiv:1712.07664}}].

\bibitem{Carr:1976zz}
B.~J. Carr, {\it {Some cosmological consequences of primordial black-hole
  evaporations}},  {\em Astrophys. J.} {\bf 206} (1976) 8--25.

\bibitem{MacGibbon:1990zk}
J.~H. MacGibbon and B.~R. Webber, {\it {Quark and gluon jet emission from
  primordial black holes: The instantaneous spectra}},  {\em Phys. Rev.} {\bf
  D41} (1990) 3052--3079.

\bibitem{MacGibbon:1991tj}
J.~H. MacGibbon, {\it {Quark and gluon jet emission from primordial black
  holes. 2. The Lifetime emission}},  {\em Phys. Rev.} {\bf D44} (1991)
  376--392.

\bibitem{Halzen:1995hu}
F.~Halzen, B.~Keszthelyi, and E.~Zas, {\it {Neutrinos from primordial black
  holes}},  {\em Phys. Rev.} {\bf D52} (1995) 3239--3247,
  [\href{http://arxiv.org/abs/hep-ph/9502268}{{\tt hep-ph/9502268}}].

\bibitem{Bugaev:2000bz}
E.~V. Bugaev and K.~V. Konishchev, {\it {Constraints on diffuse neutrino
  background from primordial black holes}},  {\em Phys. Rev.} {\bf D65} (2002)
  123005, [\href{http://arxiv.org/abs/astro-ph/0005295}{{\tt
  astro-ph/0005295}}].

\bibitem{Bugaev:2002yt}
E.~V. Bugaev and K.~V. Konishchev, {\it {Cosmological constraints from
  evaporations of primordial black holes}},  {\em Phys. Rev.} {\bf D66} (2002)
  084004, [\href{http://arxiv.org/abs/astro-ph/0206082}{{\tt
  astro-ph/0206082}}].

\bibitem{Bambeck:2005bz}
D.~Bambeck and W.~A. Hiscock, {\it {Effects of nonzero neutrino masses on black
  hole evaporation}},  {\em Class. Quant. Grav.} {\bf 22} (2005) 4247--4252,
  [\href{http://arxiv.org/abs/gr-qc/0506050}{{\tt gr-qc/0506050}}].

\bibitem{Dave:2019epr}
{\bf IceCube} Collaboration, P.~Dave and I.~Taboada, {\it {Neutrinos from
  Primordial Black Hole Evaporation}},  in {\em {HAWC Contributions to the 36th
  International Cosmic Ray Conference (ICRC2019)}}, 2019.
\newblock \href{http://arxiv.org/abs/1908.05403}{{\tt arXiv:1908.05403}}.

\bibitem{Malek:2002ns}
{\bf Super-Kamiokande} Collaboration, M.~Malek et~al., {\it {Search for
  supernova relic neutrinos at SUPER-KAMIOKANDE}},  {\em Phys. Rev. Lett.} {\bf
  90} (2003) 061101, [\href{http://arxiv.org/abs/hep-ex/0209028}{{\tt
  hep-ex/0209028}}].

\bibitem{Hooper:2019gtx}
D.~Hooper, G.~Krnjaic, and S.~D. McDermott, {\it {Dark Radiation and Superheavy
  Dark Matter from Black Hole Domination}},  {\em JHEP} {\bf 08} (2019) 001,
  [\href{http://arxiv.org/abs/1905.01301}{{\tt arXiv:1905.01301}}].

\bibitem{Kayser:1982br}
B.~Kayser, {\it {Majorana Neutrinos and their Electromagnetic Properties}},
  {\em Phys. Rev.} {\bf D26} (1982) 1662.

\bibitem{Nieves:1981zt}
J.~F. Nieves, {\it {Electromagnetic Properties of Majorana Neutrinos}},  {\em
  Phys. Rev.} {\bf D26} (1982) 3152.

\bibitem{Kayser:1983wm}
B.~Kayser and A.~S. Goldhaber, {\it {{CPT} and {CP} Properties of Majorana
  Particles, and the Consequences}},  {\em Phys. Rev.} {\bf D28} (1983) 2341.

\bibitem{Menon:2008wa}
A.~Menon and A.~M. Thalapillil, {\it {Interaction of Dirac and Majorana
  Neutrinos with Weak Gravitational Fields}},  {\em Phys. Rev.} {\bf D78}
  (2008) 113003, [\href{http://arxiv.org/abs/0804.3833}{{\tt
  arXiv:0804.3833}}].

\bibitem{Unruh:1976fm}
W.~G. Unruh, {\it {Absorption Cross-Section of Small Black Holes}},  {\em Phys.
  Rev.} {\bf D14} (1976) 3251--3259.

\bibitem{Ukwatta:2015iba}
T.~N. Ukwatta, D.~R. Stump, J.~T. Linnemann, J.~H. MacGibbon, S.~S. Marinelli,
  T.~Yapici, and K.~Tollefson, {\it {Primordial Black Holes: Observational
  Characteristics of The Final Evaporation}},  {\em Astropart. Phys.} {\bf 80}
  (2016) 90--114, [\href{http://arxiv.org/abs/1510.04372}{{\tt
  arXiv:1510.04372}}].

\bibitem{Kajita:2016cak}
T.~Kajita, {\it {Nobel Lecture: Discovery of atmospheric neutrino
  oscillations}},  {\em Rev. Mod. Phys.} {\bf 88} (2016), no.~3 030501.

\bibitem{McDonald:2016ixn}
A.~B. McDonald, {\it {Nobel Lecture: The Sudbury Neutrino Observatory:
  Observation of flavor change for solar neutrinos}},  {\em Rev. Mod. Phys.}
  {\bf 88} (2016), no.~3 030502.

\bibitem{Esteban:2018azc}
I.~Esteban, M.~C. Gonzalez-Garcia, A.~Hernandez-Cabezudo, M.~Maltoni, and
  T.~Schwetz, {\it {Global analysis of three-flavour neutrino oscillations:
  synergies and tensions in the determination of $\theta_{23}, \delta_{CP}$,
  and the mass ordering}},  {\em JHEP} {\bf 01} (2019) 106,
  [\href{http://arxiv.org/abs/1811.05487}{{\tt arXiv:1811.05487}}].

\bibitem{Page:1976df}
D.~N. Page, {\it {Particle Emission Rates from a Black Hole: Massless Particles
  from an Uncharged, Nonrotating Hole}},  {\em Phys. Rev.} {\bf D13} (1976)
  198--206.

\bibitem{Page:1976ki}
D.~N. Page, {\it {Particle Emission Rates from a Black Hole. 2. Massless
  Particles from a Rotating Hole}},  {\em Phys. Rev.} {\bf D14} (1976)
  3260--3273.

\bibitem{Page:1977um}
D.~N. Page, {\it {Particle Emission Rates from a Black Hole. 3. Charged Leptons
  from a Nonrotating Hole}},  {\em Phys. Rev.} {\bf D16} (1977) 2402--2411.

\bibitem{Giunti:2003dg}
C.~Giunti, {\it {Fock states of flavor neutrinos are unphysical}},  {\em Eur.
  Phys. J.} {\bf C39} (2005) 377--382,
  [\href{http://arxiv.org/abs/hep-ph/0312256}{{\tt hep-ph/0312256}}].

\bibitem{Ho:2012yja}
C.~M. Ho, {\it {On Neutrino Flavor States}},  {\em JHEP} {\bf 12} (2012) 022,
  [\href{http://arxiv.org/abs/1209.3453}{{\tt arXiv:1209.3453}}].

\bibitem{Shrock:1980vy}
R.~E. Shrock, {\it {New Tests For, and Bounds On, Neutrino Masses and Lepton
  Mixing}},  {\em Phys. Lett.} {\bf 96B} (1980) 159--164.

\bibitem{Shrock:1980ct}
R.~E. Shrock, {\it {General Theory of Weak Leptonic and Semileptonic Decays. 1.
  Leptonic Pseudoscalar Meson Decays, with Associated Tests For, and Bounds on,
  Neutrino Masses and Lepton Mixing}},  {\em Phys. Rev.} {\bf D24} (1981) 1232.

\bibitem{Shrock:1981wq}
R.~E. Shrock, {\it {General Theory of Weak Processes Involving Neutrinos. 2.
  Pure Leptonic Decays}},  {\em Phys. Rev.} {\bf D24} (1981) 1275.

\bibitem{Mohapatra:1979ia}
R.~N. Mohapatra and G.~Senjanovic, {\it {Neutrino Mass and Spontaneous Parity
  Violation}},  {\em Phys. Rev. Lett.} {\bf 44} (1980) 912.

\bibitem{GellMann:1980vs}
M.~Gell-Mann, P.~Ramond, and R.~Slansky, {\it {Complex Spinors and Unified
  Theories}},  {\em Conf. Proc.} {\bf C790927} (1979) 315--321,
  [\href{http://arxiv.org/abs/1306.4669}{{\tt arXiv:1306.4669}}].

\bibitem{Mohapatra:1980yp}
R.~N. Mohapatra and G.~Senjanovic, {\it {Neutrino Masses and Mixings in Gauge
  Models with Spontaneous Parity Violation}},  {\em Phys. Rev.} {\bf D23}
  (1981) 165.

\bibitem{Schechter:1980gr}
J.~Schechter and J.~W.~F. Valle, {\it {Neutrino Masses in SU(2) x U(1)
  Theories}},  {\em Phys. Rev.} {\bf D22} (1980) 2227.

\bibitem{Lazarides:1980nt}
G.~Lazarides, Q.~Shafi, and C.~Wetterich, {\it {Proton Lifetime and Fermion
  Masses in an SO(10) Model}},  {\em Nucl. Phys.} {\bf B181} (1981) 287--300.

\bibitem{Mohapatra:1986bd}
R.~N. Mohapatra and J.~W.~F. Valle, {\it {Neutrino Mass and Baryon Number
  Nonconservation in Superstring Models}},  {\em Phys. Rev.} {\bf D34} (1986)
  1642.

\bibitem{Foot:1988aq}
R.~Foot, H.~Lew, X.~G. He, and G.~C. Joshi, {\it {Seesaw Neutrino Masses
  Induced by a Triplet of Leptons}},  {\em Z. Phys.} {\bf C44} (1989) 441.

\bibitem{Balantekin:2018azf}
A.~Baha~Balantekin and B.~Kayser, {\it {On the Properties of Neutrinos}},  {\em
  Ann. Rev. Nucl. Part. Sci.} {\bf 68} (2018) 313--338,
  [\href{http://arxiv.org/abs/1805.00922}{{\tt arXiv:1805.00922}}].

\bibitem{Traschen:1999zr}
J.~H. Traschen, {\it {An Introduction to black hole evaporation}},  in {\em
  {Mathematical methods in physics. Proceedings, Winter School, Londrina,
  Brazil, August 17-26, 1999}}, 1999.
\newblock \href{http://arxiv.org/abs/gr-qc/0010055}{{\tt gr-qc/0010055}}.

\bibitem{birrell_davies_1982}
N.~D. Birrell and P.~C.~W. Davies, {\em Quantum Fields in Curved Space}.
\newblock Cambridge Monographs on Mathematical Physics. Cambridge University
  Press, 1982.

\bibitem{Hannestad:2010kz}
S.~Hannestad, {\it {Neutrino physics from precision cosmology}},  {\em Prog.
  Part. Nucl. Phys.} {\bf 65} (2010) 185--208,
  [\href{http://arxiv.org/abs/1007.0658}{{\tt arXiv:1007.0658}}].

\bibitem{Diaz:2015aua}
J.~Diaz and F.~Klinkhamer, {\it {Neutrino refraction by the cosmic neutrino
  background}},  {\em Phys. Rev. D} {\bf 93} (2016), no.~5 053004,
  [\href{http://arxiv.org/abs/1512.00817}{{\tt arXiv:1512.00817}}].

\bibitem{Fujita:2014hha}
T.~Fujita, M.~Kawasaki, K.~Harigaya, and R.~Matsuda, {\it {Baryon asymmetry,
  dark matter, and density perturbation from primordial black holes}},  {\em
  Phys. Rev.} {\bf D89} (2014), no.~10 103501,
  [\href{http://arxiv.org/abs/1401.1909}{{\tt arXiv:1401.1909}}].

\bibitem{Hamada:2016jnq}
Y.~Hamada and S.~Iso, {\it {Baryon asymmetry from primordial black holes}},
  {\em PTEP} {\bf 2017} (2017), no.~3 033B02,
  [\href{http://arxiv.org/abs/1610.02586}{{\tt arXiv:1610.02586}}].

\bibitem{Morrison:2018xla}
L.~Morrison, S.~Profumo, and Y.~Yu, {\it {Melanopogenesis: Dark Matter of
  (almost) any Mass and Baryonic Matter from the Evaporation of Primordial
  Black Holes weighing a Ton (or less)}},  {\em JCAP} {\bf 1905} (2019), no.~05
  005, [\href{http://arxiv.org/abs/1812.10606}{{\tt arXiv:1812.10606}}].

\bibitem{Doran:2005vm}
C.~Doran, A.~Lasenby, S.~Dolan, and I.~Hinder, {\it {Fermion absorption cross
  section of a Schwarzschild black hole}},  {\em Phys. Rev.} {\bf D71} (2005)
  124020, [\href{http://arxiv.org/abs/gr-qc/0503019}{{\tt gr-qc/0503019}}].

\bibitem{Chen:2015dka}
M.-C. Chen, M.~Ratz, and A.~Trautner, {\it {Nonthermal cosmic neutrino
  background}},  {\em Phys. Rev.} {\bf D92} (2015), no.~12 123006,
  [\href{http://arxiv.org/abs/1509.00481}{{\tt arXiv:1509.00481}}].

\bibitem{Antonelli:1981eg}
F.~Antonelli, D.~Fargion, and R.~Konoplich, {\it {Right-handed Neutrino
  Interactions in the Early Universe}},  {\em Lett. Nuovo Cim.} {\bf 32} (1981)
  289.

\bibitem{Aghanim:2018eyx}
{\bf Planck} Collaboration, N.~Aghanim et~al., {\it {Planck 2018 results. VI.
  Cosmological parameters}},  \href{http://arxiv.org/abs/1807.06209}{{\tt
  arXiv:1807.06209}}.

\bibitem{Abazajian:2019oqj}
K.~N. Abazajian and J.~Heeck, {\it {Observing Dirac neutrinos in the cosmic
  microwave background}},  \href{http://arxiv.org/abs/1908.03286}{{\tt
  arXiv:1908.03286}}.

\bibitem{Benson:2014qhw}
{\bf SPT-3G} Collaboration, B.~A. Benson et~al., {\it {SPT-3G: A
  Next-Generation Cosmic Microwave Background Polarization Experiment on the
  South Pole Telescope}},  {\em Proc. SPIE Int. Soc. Opt. Eng.} {\bf 9153}
  (2014) 91531P, [\href{http://arxiv.org/abs/1407.2973}{{\tt
  arXiv:1407.2973}}].

\bibitem{Abitbol:2019nhf}
{\bf Simons Observatory} Collaboration, M.~H. Abitbol et~al., {\it {The Simons
  Observatory: Astro2020 Decadal Project Whitepaper}},
  \href{http://arxiv.org/abs/1907.08284}{{\tt arXiv:1907.08284}}.

\bibitem{Abazajian:2019eic}
K.~Abazajian et~al., {\it {CMB-S4 Science Case, Reference Design, and Project
  Plan}},  \href{http://arxiv.org/abs/1907.04473}{{\tt arXiv:1907.04473}}.

\bibitem{Carr:1975qj}
B.~J. Carr, {\it {The Primordial black hole mass spectrum}},  {\em Astrophys.
  J.} {\bf 201} (1975) 1--19.

\bibitem{Green:1999yh}
A.~M. Green, {\it {Supersymmetry and primordial black hole abundance
  constraints}},  {\em Phys. Rev.} {\bf D60} (1999) 063516,
  [\href{http://arxiv.org/abs/astro-ph/9903484}{{\tt astro-ph/9903484}}].

\bibitem{Zeldovich:1977aa}
Y.~B. Zel'dovich, A.~A. Starobinskii, M.~I. Khlopov, and V.~M. Chechetkin, {\it
  {Primordial black holes and the deuterium problem}},  {\em Pis'ma Astron.
  Zh.} {\bf 3} (1977) 208. [Sov. Astron. Lett. 3 110 (1977)].

\bibitem{MacGibbon:1987my}
J.~H. MacGibbon, {\it {Can Planck-mass relics of evaporating black holes close
  the universe?}},  {\em Nature} {\bf 329} (1987) 308--309.

\bibitem{Barrow:1992hq}
J.~D. Barrow, E.~J. Copeland, and A.~R. Liddle, {\it {The Cosmology of black
  hole relics}},  {\em Phys. Rev.} {\bf D46} (1992) 645--657.

\bibitem{Carr:1994ar}
B.~J. Carr, J.~H. Gilbert, and J.~E. Lidsey, {\it {Black hole relics and
  inflation: Limits on blue perturbation spectra}},  {\em Phys. Rev.} {\bf D50}
  (1994) 4853--4867, [\href{http://arxiv.org/abs/astro-ph/9405027}{{\tt
  astro-ph/9405027}}].

\bibitem{Nozari:2005ah}
K.~Nozari and S.~H. Mehdipour, {\it {Gravitational uncertainty and black hole
  remnants}},  {\em Mod. Phys. Lett.} {\bf A20} (2005) 2937--2948,
  [\href{http://arxiv.org/abs/0809.3144}{{\tt arXiv:0809.3144}}].

\bibitem{Chen:2004ft}
P.~Chen, {\it {Inflation induced Planck-size black hole remnants as dark
  matter}},  {\em New Astron. Rev.} {\bf 49} (2005) 233--239,
  [\href{http://arxiv.org/abs/astro-ph/0406514}{{\tt astro-ph/0406514}}].

\bibitem{Barrau:2003xp}
A.~Barrau, D.~Blais, G.~Boudoul, and D.~Polarski, {\it {Peculiar relics from
  primordial black holes in the inflationary paradigm}},  {\em Annalen Phys.}
  {\bf 13} (2004) 115--123, [\href{http://arxiv.org/abs/astro-ph/0303330}{{\tt
  astro-ph/0303330}}].

\bibitem{Chen:2002tu}
P.~Chen and R.~J. Adler, {\it {Black hole remnants and dark matter}},  {\em
  Nucl. Phys. Proc. Suppl.} {\bf 124} (2003) 103--106,
  [\href{http://arxiv.org/abs/gr-qc/0205106}{{\tt gr-qc/0205106}}].
  [,103(2002)].

\bibitem{Alexeyev:2002tg}
S.~Alexeyev, A.~Barrau, G.~Boudoul, O.~Khovanskaya, and M.~Sazhin, {\it {Black
  hole relics in string gravity: Last stages of Hawking evaporation}},  {\em
  Class. Quant. Grav.} {\bf 19} (2002) 4431--4444,
  [\href{http://arxiv.org/abs/gr-qc/0201069}{{\tt gr-qc/0201069}}].

\bibitem{Susskind:1995da}
L.~Susskind, {\it {Trouble for remnants}},
  \href{http://arxiv.org/abs/hep-th/9501106}{{\tt hep-th/9501106}}.

\bibitem{Barrow:1991dn}
J.~D. Barrow, E.~J. Copeland, and A.~R. Liddle, {\it {The Evolution of black
  holes in an expanding universe}},  {\em Mon. Not. Roy. Astron. Soc.} {\bf
  253} (1991) 675--682.

\bibitem{Bernal:2019lpc}
N.~Bernal and F.~Hajkarim, {\it {Primordial Gravitational Waves in Non-standard
  Cosmologies}},  \href{http://arxiv.org/abs/1905.10410}{{\tt
  arXiv:1905.10410}}.

\bibitem{Arias:2019uol}
P.~Arias, N.~Bernal, A.~Herrera, and C.~Maldonado, {\it {Reconstructing
  Non-standard Cosmologies with Dark Matter}},
  \href{http://arxiv.org/abs/1906.04183}{{\tt arXiv:1906.04183}}.

\bibitem{deSalas:2016ztq}
P.~F. de~Salas and S.~Pastor, {\it {Relic neutrino decoupling with flavour
  oscillations revisited}},  {\em JCAP} {\bf 1607} (2016), no.~07 051,
  [\href{http://arxiv.org/abs/1606.06986}{{\tt arXiv:1606.06986}}].

\bibitem{Riess:2016jrr}
A.~G. Riess et~al., {\it {A 2.4\% Determination of the Local Value of the
  Hubble Constant}},  {\em Astrophys. J.} {\bf 826} (2016), no.~1 56,
  [\href{http://arxiv.org/abs/1604.01424}{{\tt arXiv:1604.01424}}].

\bibitem{Escudero:2019gzq}
M.~Escudero, D.~Hooper, G.~Krnjaic, and M.~Pierre, {\it {Cosmology With a Very
  Light $L_\mu - L_\tau$ Gauge Boson}},
  \href{http://arxiv.org/abs/1901.02010}{{\tt arXiv:1901.02010}}.

\bibitem{Vagnozzi:2019ezj}
S.~Vagnozzi, {\it {New physics in light of the $H_0$ tension: an alternative
  view}},  \href{http://arxiv.org/abs/1907.07569}{{\tt arXiv:1907.07569}}.

\bibitem{Katz:2018zrn}
A.~Katz, J.~Kopp, S.~Sibiryakov, and W.~Xue, {\it {Femtolensing by Dark Matter
  Revisited}},  {\em JCAP} {\bf 1812} (2018) 005,
  [\href{http://arxiv.org/abs/1807.11495}{{\tt arXiv:1807.11495}}].

\bibitem{Kim:1999iv}
H.~I. Kim, C.~H. Lee, and J.~H. MacGibbon, {\it {Diffuse gamma-ray background
  and primordial black hole constraints on the spectral index of density
  fluctuations}},  {\em Phys. Rev.} {\bf D59} (1999) 063004,
  [\href{http://arxiv.org/abs/astro-ph/9901030}{{\tt astro-ph/9901030}}].

\bibitem{Bahcall:2000nu}
J.~N. Bahcall, M.~H. Pinsonneault, and S.~Basu, {\it {Solar models: Current
  epoch and time dependences, neutrinos, and helioseismological properties}},
  {\em Astrophys. J.} {\bf 555} (2001) 990--1012,
  [\href{http://arxiv.org/abs/astro-ph/0010346}{{\tt astro-ph/0010346}}].

\bibitem{Bahcall:2005va}
J.~N. Bahcall, A.~M. Serenelli, and S.~Basu, {\it {10,000 standard solar
  models: a Monte Carlo simulation}},  {\em Astrophys. J. Suppl.} {\bf 165}
  (2006) 400--431, [\href{http://arxiv.org/abs/astro-ph/0511337}{{\tt
  astro-ph/0511337}}].

\bibitem{Vitagliano:2017odj}
E.~Vitagliano, J.~Redondo, and G.~Raffelt, {\it {Solar neutrino flux at keV
  energies}},  {\em JCAP} {\bf 1712} (2017), no.~12 010,
  [\href{http://arxiv.org/abs/1708.02248}{{\tt arXiv:1708.02248}}].

\bibitem{Battistoni:2005pd}
G.~Battistoni, A.~Ferrari, T.~Montaruli, and P.~R. Sala, {\it {The atmospheric
  neutrino flux below 100-MeV: The FLUKA results}},  {\em Astropart. Phys.}
  {\bf 23} (2005) 526--534.

\bibitem{Honda:2011nf}
M.~Honda, T.~Kajita, K.~Kasahara, and S.~Midorikawa, {\it {Improvement of low
  energy atmospheric neutrino flux calculation using the JAM nuclear
  interaction model}},  {\em Phys. Rev.} {\bf D83} (2011) 123001,
  [\href{http://arxiv.org/abs/1102.2688}{{\tt arXiv:1102.2688}}].

\bibitem{Beacom:2010kk}
J.~F. Beacom, {\it {The Diffuse Supernova Neutrino Background}},  {\em Ann.
  Rev. Nucl. Part. Sci.} {\bf 60} (2010) 439--462,
  [\href{http://arxiv.org/abs/1004.3311}{{\tt arXiv:1004.3311}}].

\bibitem{Dolgov:2002wy}
A.~D. Dolgov, {\it {Neutrinos in cosmology}},  {\em Phys. Rept.} {\bf 370}
  (2002) 333--535, [\href{http://arxiv.org/abs/hep-ph/0202122}{{\tt
  hep-ph/0202122}}].

\bibitem{Dolgov:2008hz}
A.~D. Dolgov, {\it {Cosmology and Neutrino Properties}},  {\em Phys. Atom.
  Nucl.} {\bf 71} (2008) 2152--2164,
  [\href{http://arxiv.org/abs/0803.3887}{{\tt arXiv:0803.3887}}].

\bibitem{Lesgourgues:2006nd}
J.~Lesgourgues and S.~Pastor, {\it {Massive neutrinos and cosmology}},  {\em
  Phys. Rept.} {\bf 429} (2006) 307--379,
  [\href{http://arxiv.org/abs/astro-ph/0603494}{{\tt astro-ph/0603494}}].

\bibitem{Quigg:2008ab}
C.~Quigg, {\it {Cosmic Neutrinos}},  in {\em {Proceedings, 35th SLAC Summer
  Institute on Particle Physics: Dark matter: From the cosmos to the Laboratory
  (SSI 2007): Menlo Park, California, July 30- August 10, 2007}}, 2008.
\newblock \href{http://arxiv.org/abs/0802.0013}{{\tt arXiv:0802.0013}}.

\bibitem{Ivanchik:2018fxy}
A.~V. Ivanchik and V.~{\relax Yu}. Yurchenko, {\it {Relic neutrinos:
  Antineutrinos of Primordial Nucleosynthesis}},  {\em Phys. Rev.} {\bf D98}
  (2018), no.~8 081301, [\href{http://arxiv.org/abs/1809.03349}{{\tt
  arXiv:1809.03349}}].

\bibitem{Betti:2019ouf}
{\bf PTOLEMY} Collaboration, M.~G. Betti et~al., {\it {Neutrino Physics with
  the PTOLEMY project}},  \href{http://arxiv.org/abs/1902.05508}{{\tt
  arXiv:1902.05508}}.

\bibitem{Long:2014zva}
A.~J. Long, C.~Lunardini, and E.~Sabancilar, {\it {Detecting non-relativistic
  cosmic neutrinos by capture on tritium: phenomenology and physics
  potential}},  {\em JCAP} {\bf 1408} (2014) 038,
  [\href{http://arxiv.org/abs/1405.7654}{{\tt arXiv:1405.7654}}].

\bibitem{McKeen:2018xyz}
D.~McKeen, {\it {Cosmic neutrino background search experiments as decaying dark
  matter detectors}},  {\em Phys. Rev.} {\bf D100} (2019), no.~1 015028,
  [\href{http://arxiv.org/abs/1812.08178}{{\tt arXiv:1812.08178}}].

\bibitem{Chacko:2018uke}
Z.~Chacko, P.~Du, and M.~Geller, {\it {Detecting a Secondary Cosmic Neutrino
  Background from Majoron Decays in Neutrino Capture Experiments}},  {\em Phys.
  Rev.} {\bf D100} (2019), no.~1 015050,
  [\href{http://arxiv.org/abs/1812.11154}{{\tt arXiv:1812.11154}}].

\bibitem{Weinberg:1962zza}
S.~Weinberg, {\it {Universal Neutrino Degeneracy}},  {\em Phys. Rev.} {\bf 128}
  (1962) 1457--1473.

\bibitem{Cocco:2007za}
A.~G. Cocco, G.~Mangano, and M.~Messina, {\it {Probing low energy neutrino
  backgrounds with neutrino capture on beta decaying nuclei}},  {\em JCAP} {\bf
  0706} (2007) 015, [\href{http://arxiv.org/abs/hep-ph/0703075}{{\tt
  hep-ph/0703075}}].

\bibitem{Cocco:2009rh}
A.~G. Cocco, G.~Mangano, and M.~Messina, {\it {Low Energy Antineutrino
  Detection Using Neutrino Capture on EC Decaying Nuclei}},  {\em Phys. Rev.}
  {\bf D79} (2009) 053009, [\href{http://arxiv.org/abs/0903.1217}{{\tt
  arXiv:0903.1217}}].

\end{thebibliography}\endgroup

\end{document}